\documentclass[twocolumn]{aastex62}
\usepackage{xcolor,soul}
\providecommand{\bjdtdb}{\ensuremath{\rm {BJD_{TDB}}}}

\providecommand{\msun}{\ensuremath{\,M_\Sun}}
\providecommand{\rsun}{\ensuremath{\,R_\Sun}}
\providecommand{\lsun}{\ensuremath{\,L_\Sun}}
\providecommand{\mj}{\ensuremath{\,{\rm M_J}}}
\providecommand{\rj}{\ensuremath{\,{\rm R_J}}}

\providecommand{\mst}{\ensuremath{\,{\rm M_\odot}}}
\providecommand{\rst}{\ensuremath{\,{\rm R_\odot}}}

\providecommand{\arcsec}{$^{\prime \prime}$}

\graphicspath{{./}{figures/}}

\providecommand{\dbf}{}
\newcommand\dsout[1]{}

\begin{document}

\title{TOI-811b and TOI-852b: New transiting brown dwarfs with similar masses and very different radii and ages from the TESS mission}

\author[0000-0001-6416-1274]{Theron W. Carmichael}
\affil{\rm Harvard University,
Cambridge, MA 02138}
\affil{\rm Center for Astrophysics ${\rm \mid}$ Harvard {\rm \&} Smithsonian, 60 Garden Street, Cambridge, MA 02138, USA}
\affil{\rm National Science Foundation Graduate Research Fellow}

\author{Samuel N. Quinn}
\affil{\rm Center for Astrophysics ${\rm \mid}$ Harvard {\rm \&} Smithsonian, 60 Garden Street, Cambridge, MA 02138, USA}

\author{George Zhou}
\affil{\rm Center for Astrophysics ${\rm \mid}$ Harvard {\rm \&} Smithsonian, 60 Garden Street, Cambridge, MA 02138, USA}

\author{Nolan Grieves}
\affil{\rm Geneva Observatory, University of Geneva, Chemin des Mailettes 51, 1290 Versoix, Switzerland}

\author{Jonathan M. Irwin}
\affil{\rm Center for Astrophysics ${\rm \mid}$ Harvard {\rm \&} Smithsonian, 60 Garden Street, Cambridge, MA 02138, USA}

\author[0000-0002-3481-9052]{Keivan G.\ Stassun}
\affil{\rm Vanderbilt University, Department of Physics \& Astronomy, 6301 Stevenson Center Ln., Nashville, TN 37235, USA}
\affil{\rm Fisk University, Department of Physics, 1000 18th Ave. N., Nashville, TN 37208, USA}

\author{Andrew M. Vanderburg}
\affil{\rm Department of Astronomy, University of Wisconsin-Madison, Madison, WI 53706, USA}

\author[0000-0002-4265-047X]{Joshua N. Winn}
\affil{\rm Department of Astrophysical Sciences, Princeton University, 4 Ivy Lane, Princeton, NJ 08544, USA}

\author{Francois Bouchy}
\affil{\rm Geneva Observatory, University of Geneva, Chemin des Mailettes 51, 1290 Versoix, Switzerland} 

\author[0000-0002-9314-960X]{Clara E. Brasseur}
\affil{\rm Space Telescope Science Institute, USA}

\author{C\'{e}sar Brice\~{n}o}
\affiliation{\rm Cerro Tololo Inter-American Observatory, Casilla 603, La Serena, Chile} 

\author[0000-0003-1963-9616]{Douglas A. Caldwell}
\affil{\rm SETI Institute/NASA Ames Research Center, Moffett Field, CA 94035, USA}

\author{David Charbonneau}
\affil{\rm Center for Astrophysics ${\rm \mid}$ Harvard {\rm \&} Smithsonian, 60 Garden Street, Cambridge, MA 02138, USA}

\author[0000-0001-6588-9574]{Karen A.\ Collins} 
\affiliation{\rm Center for Astrophysics ${\rm \mid}$ Harvard {\rm \&} Smithsonian, 60 Garden Street, Cambridge, MA 02138, USA}

\author{Knicole D. Colon}
\affiliation{\rm NASA Goddard Space Flight Center, Exoplanets and Stellar Astrophysics Laboratory (Code 667), Greenbelt, MD 20771, USA}

\author{Jason D. Eastman}
\affiliation{\rm Center for Astrophysics ${\rm \mid}$ Harvard {\rm \&} Smithsonian, 60 Garden Street, Cambridge, MA 02138, USA}

\author[0000-0002-9113-7162]{Michael Fausnaugh}
\affil{\rm Department of Physics, and Kavli Institute for Astrophysics and Space Research, Massachusetts Institute of Technology, Cambridge, MA 02139, USA}

\author{William Fong}
\affil{\rm Department of Physics, and Kavli Institute for Astrophysics and Space Research, Massachusetts Institute of Technology, Cambridge, MA 02139, USA}

\author{G{\'a}bor~F{\H u}r{\'e}sz}
\affil{\rm Department of Physics, and Kavli Institute for Astrophysics and Space Research, Massachusetts Institute of Technology, Cambridge, MA 02139, USA}

\author{Chelsea Huang}
\affil{\rm Department of Physics, and Kavli Institute for Astrophysics and Space Research, Massachusetts Institute of Technology, Cambridge, MA 02139, USA}

\author[0000-0002-4715-9460]{Jon M. Jenkins}
\affil{\rm NASA Ames Research Center, Moffett Field, CA 94035, USA}

\author[0000-0003-0497-2651]{John F.\ Kielkopf} 
\affiliation{\rm Department of Physics and Astronomy, University of Louisville, Louisville, KY 40292, USA}

\author{David W. Latham}
\affil{\rm Center for Astrophysics ${\rm \mid}$ Harvard {\rm \&} Smithsonian, 60 Garden Street, Cambridge, MA 02138, USA}

\author{Nicholas Law}
\affiliation{\rm Department of Physics and Astronomy, The University of North Carolina at Chapel Hill, Chapel Hill, NC 27599-3255, USA}

\author{Michael B. Lund}
\affiliation{\rm IPAC-NASA Exoplanet Science Institute, Pasadena, CA 91125, USA}

\author[0000-0003-3654-1602]{Andrew W. Mann}
\affiliation{\rm Department of Physics and Astronomy, The University of North Carolina at Chapel Hill, Chapel Hill, NC 27599-3255, USA}

\author{George R. Ricker}
\affil{\rm Department of Physics, and Kavli Institute for Astrophysics and Space Research, Massachusetts Institute of Technology, Cambridge, MA 02139, USA}

\author{Joseph E. Rodriguez}
\affil{\rm Michigan State University, East Lansing, MI, 48824, USA}
\affil{\rm Center for Astrophysics ${\rm \mid}$ Harvard {\rm \&} Smithsonian, 60 Garden Street, Cambridge, MA 02138, USA}

\author[0000-0001-8227-1020]{Richard P. Schwarz}
\affiliation{\rm Patashnick Voorheesville Observatory, Voorheesville, NY 12186, USA}

\author{Avi Shporer}
\affil{\rm Department of Physics, and Kavli Institute for Astrophysics and Space Research, Massachusetts Institute of Technology, Cambridge, MA 02139, USA}

\author[0000-0002-1949-4720]{Peter Tenenbaum}
\affil{\rm SETI Institute/NASA Ames Research Center, Moffett Field, CA 94035, USA}

\author[0000-0001-7336-7725]{Mackenna L. Wood} 
\affil{\rm Department of Physics and Astronomy, The University of North Carolina at Chapel Hill, Chapel Hill, NC 27599, USA}

\author{Carl Ziegler}
\affil{\rm Dunlap Institute for Astronomy and Astrophysics, University of Toronto, 50 St. George Street, Toronto, Ontario M5S 3H4, Canada}

\begin{abstract}
\noindent We report the discovery of two transiting brown dwarfs (BDs), TOI-811b and TOI-852b, from NASA's Transiting Exoplanet Survey Satellite mission. These two transiting BDs have similar masses, but very different radii and ages. Their host stars have similar masses, effective temperatures, and metallicities. The younger and larger transiting BD is TOI-811b at a mass of $M_b = 59.9 \pm 13.0{\rm M_J}$ and radius of $R_b = 1.26 \pm 0.06{\rm R_J}$ and it orbits its host star in a period of $P = 25.16551 \pm 0.00004$ days. \dbf{We derive the host star's age of $93^{+61}_{-29}$ Myr from an application of gyrochronology. The youth of this system, rather than external heating from its host star, is why this BD's radius is relatively large.} This constraint on the youth of TOI-811b allows us to test substellar mass-radius evolutionary models at young ages where the radius of BDs changes rapidly with age. \dbf{TOI-852b has a similar mass at  $M_b = 53.7 \pm 1.4{\rm M_J}$, but is much older (\dbf{4 or 8 Gyr, based on bimodal isochrone results of the host star}), and is also smaller with a radius of $R_b = 0.83 \pm 0.04{\rm R_J}$. TOI-852b's orbital period is $P = 4.94561 \pm 0.00008$ days.} TOI-852b joins the likes of other old transiting BDs that trace out the oldest substellar mass-radius evolutionary models where contraction of the BD's radius slows and approaches a constant value. Both host stars have a mass of $M_\star = 1.32{\rm M_\odot}\pm0.05$ and differ in their radii, $T_{\rm eff}$, and [Fe/H] with TOI-811 having $R_\star=1.27\pm0.09{\rm R_\odot}$, $T_{\rm eff} = 6107 \pm 77$K, and $\rm [Fe/H] = +0.40 \pm 0.09$ and TOI-852 having $R_\star=1.71\pm0.04{\rm R_\odot}$, $T_{\rm eff} = 5768 \pm 84$K, and $\rm [Fe/H] = +0.33 \pm 0.09$. We take this opportunity to examine how TOI-811b and TOI-852b serve as test points for young and old substellar isochrones, respectively.

\end{abstract}


\section{Introduction} \label{sec:intro}

Discoveries of transiting brown dwarfs (BDs) have become more frequent over the past two years and this has granted astronomers new opportunities to understand these objects in greater detail. Currently, the defining feature of BDs is an arbitrary mass criterion created to distinguish BDs from giant planets and stars: a mass range that is between the two loose boundaries $11-16\mj$ \citep{spiegel2011} and $75-80\mj$ \citep{baraffe2002}. The lower boundary serves to distinguish BDs from giant planets as this range of masses is the threshold in which deuterium may be fused at some point during the lifetime of the BD. The upper boundary serves to mark the threshold at which hydrogen fusion is sustainable and the object is classified as a star. This traditional definition for BDs as ``the objects between planets and stars" does not \dbf{respect the importance of processes} more fundamental than deuterium or hydrogen fusion that would be better used to define BDs. It is more intuitive to trace the definition of BDs not to what they do or do not fuse within their cores, but instead to the processes responsible for their formation \dbf{\citep[an idea explored in][for example]{burrows01}}. However, in order to explore this idea, we must first settle two important issues: 1) the scarcity of known transiting BDs available for study and 2) the need to test the substellar evolutionary models that characterize transiting BDs.

To address the scarcity of known transiting BDs (known as the brown dwarf desert), we focus our efforts on using NASA's Transiting Exoplanet Survey Satellite (TESS) mission in combination with ground-based follow up observations and data from the Gaia mission to search for and characterize new transiting BDs. The TESS mission is an all-sky survey that has successfully completed its two-year primary mission and has begun its first extended mission. TESS has been the driving force behind the discovery of new transiting BDs over the past two years \citep[for a list, see][]{carmichael20}. Light curves from the TESS mission provide us with the orbital period, orbital inclination, and a rough estimate for the radius of transiting BDs. However, the radius estimate from the TESS light curve alone is not sufficient as this only provides the ratio of the size of the BD to the size of the host star (known as the transit depth). To make effective use of this transit depth, we use stellar parallax measurements from Gaia Data Release 2 (Gaia DR2). These parallax measurements may be translated into a distance, which when combined with the spectral energy distribution (SED) of a star, give us an estimate of the \dbf{stellar luminosity and effective temperature}. \dbf{These can then be turned into an estimate for the stellar radius. With the addition of a spectroscopic surface gravity, we also have the means to estimate the mass of the host star.} Gaia DR2 is particularly special in this regard as it has increased the precision to which we are able to determine stellar \dsout{masses and}radii and made these parameters less of a limiting source of uncertainty as they were prior to Gaia DR2.

With a stellar mass and radius in hand, we may now determine the same parameters for the transiting BD. The orbital inclination of the transiting BD that we obtain from a light curve is used to determine the mass of the BD. The inclination $i$ breaks the well-known minimum mass $m\sin{i}$ degeneracy that we are limited to when only follow up radial velocity (RV) data are used. This, in combination with the semi-amplitude measured from the RVs (which is linked directly to the host star's mass estimated from Gaia DR2 and SED models), gives us the mass of the BD. To determine the radius, we use the transit depth from the light curve and the stellar radius we determined from an SED informed by broadband photometry and data from Gaia DR2 to solve for the transiting BD's radius. This procedure of using transit light curves in combination with RV follow up to measure a mass and radius is well-established and we are fortunate in this era of precise parallax measurements to be able to augment these traditional techniques further with Gaia DR2. 

It is important that we use these mass and radius measurements of transiting BDs to test substellar evolutionary models more rigorously. To test these models, we examine how well they agree with the measured masses, radii, and ages of transiting BDs. Age is an important feature---perhaps the most important---as the radius of a transiting BD is not constant for a given mass over the lifetime of the object. The radius of the BD contracts most quickly at young ages up to 1 Gyr and asymptotically decelerates in this contraction between 3 and 10 Gyr \citep{baraffe03, saumon08, burrows2011, ATMO2020}.  In detail, the substellar evolutionary models that \cite{ATMO2020} (ATMO 2020) have developed are substellar isochrone tracks where any particular age track dictates the radius of the BD. \dbf{Other works like \cite{chabrier2000}, \cite{baraffe2002}, \cite{baraffe03}, and \cite{saumon08} also employ related but distinct techniques to trace the evolution of substellar objects and giant planets.} However, the age of a transiting BD is notoriously much more difficult to obtain than either the mass or radius. Only a handful of transiting BDs \citep{ad3116, cww89a, david19_bd} have ages with low uncertainties because they are determined through star cluster membership. One of the youngest of these is AD 3116b at roughly 600 Myr of age in the Praesepe cluster. We are particularly interested in finding more transiting BDs as young or younger than this given how quickly the radius changes during this part of the BD's evolution. This means that we always seek out systems where we can measure the age of the host star and transiting BD via means other than fitting to a stellar isochrone alone. This is where we may look to gyrochronology for the first time as an application to stars that host transiting BDs.

Gyrochronology is the study of the age of solar analog stars through their colors and rotation rates \citep{mamajek_hillenbrand_2008}. This builds on what is known as the Skumanich relation \citep{skumanich_1972}, which notes that the rotation rate of solar analog stars slows down as they age. This age-rotation relationship is a powerful tool to measure the ages of transiting BDs because it means that we may rely less on finding them in stellar clusters and associations to obtain a relatively precise age estimate. Gyrochronology is also better suited to determining the ages of young solar analogs (as opposed to older stars) through the relationship between the stellar rotation rate and color, which makes this particularly useful in an application to young transiting BDs that orbit such young Sun-like stars.

Here we present the discovery of two new transiting BDs from the TESS mission. These are TOI-811b and TOI-852b and they are two transiting BDs of similar masses with very different radii and ages. TOI-811b is roughly 90 Myr old and is in a relatively wide, eccentric orbit around its host star. In contrast, TOI-852b is at least 4 Gyr old in a close, circular orbit around its host star. In Section \ref{sec:observations}, we give details on the light curves, spectra, and other observations that were obtained for each system. Section \ref{sec:analysis} describes the analysis techniques used to derive the host star and BD properties. Section \ref{sec:conclusion} contains discussion of the implications of these new discoveries in the BD mass-radius diagram with particular focus on TOI-811b, which is one of the youngest known transiting BDs with a well-determined radius.

\section{Observations}\label{sec:observations}
\subsection{TESS and ground-based light curves}
The initial detections of the transits of both BDs were made by the TESS mission. \dbf{We provide a summary of the facilities used in this work in Table \ref{tab:telescopes}} and give a list of equatorial coordinates and magnitudes in Table \ref{tab:toi_obs}. TOI-811 \dbf{(TIC 100757807)} was observed in Sector 3 at 30-minute cadence and TOI-852 \dbf{(TIC 29918916)} was observed in Sectors 5 and 6 also at a 30-minute cadence. These transits were detected in the MIT Quick Look Pipeline (QLP, C. X. Huang et al. 2020, in preparation). Both BDs received additional ground-based, seeing-limited photometric follow up to aid in the confirmation that the transits originate from the target stars. TOI-811 was observed by the MEarth array of telescopes on Mt. Hopkins, Arizona, the Las Cumbres Observatory (LCOGT) 1-m telescope, and the 0.36-m telescope at El Sauce Observatory in Coquimbo Province, Chile. TOI-852 was observed by the University of Louisville Manner Telescope (ULMT) on Mt. Lemmon.

\begin{deluxetable}{lcc}
 \tabletypesize{\footnotesize}
 \tablewidth{0pt}
 \tablecaption{List of facilities used for each system. This table serves as a reference to which facilities we use to confirm and analyze the BDs in each system. The data from El Sauce, LCOGT, and WASP are not used in the transit photometry analysis in {\tt EXOFASTv2}. \label{tab:telescopes}}
 \tablehead{
 \colhead{Facility} & \colhead{Data type} & \colhead{Object}}
 \startdata
 TESS & Transit photometry & TOI-811, TOI-852\\
 MEarth & Transit photometry & TOI-811 \\
 LCOGT & Transit photometry & TOI-811 \\
 El Sauce & Transit photometry & TOI-811 \\
 ULMT & Transit photometry & TOI-852 \\
 WASP & Rotational light curve & TOI-811, TOI-852 \\
 SOAR & Speckle imaging & TOI-811, TOI-852 \\
 Gaia & Parallax \& proper motion & TOI-811, TOI-852 \\
 CHIRON & Spectroscopy \& RVs & TOI-811 \\
 TRES & Spectroscopy \& RVs & TOI-811, TOI-852 \\
 CORALIE & Spectroscopy \& RVs & TOI-852 \\
 \hline
 \enddata
\end{deluxetable}

\subsubsection{TESS light curves}\label{sec:tess_lc}
\begin{figure}[!ht]
\centering
\includegraphics[width=0.49\textwidth, trim={1.0cm 0.0cm 0.0cm 0.0cm}]{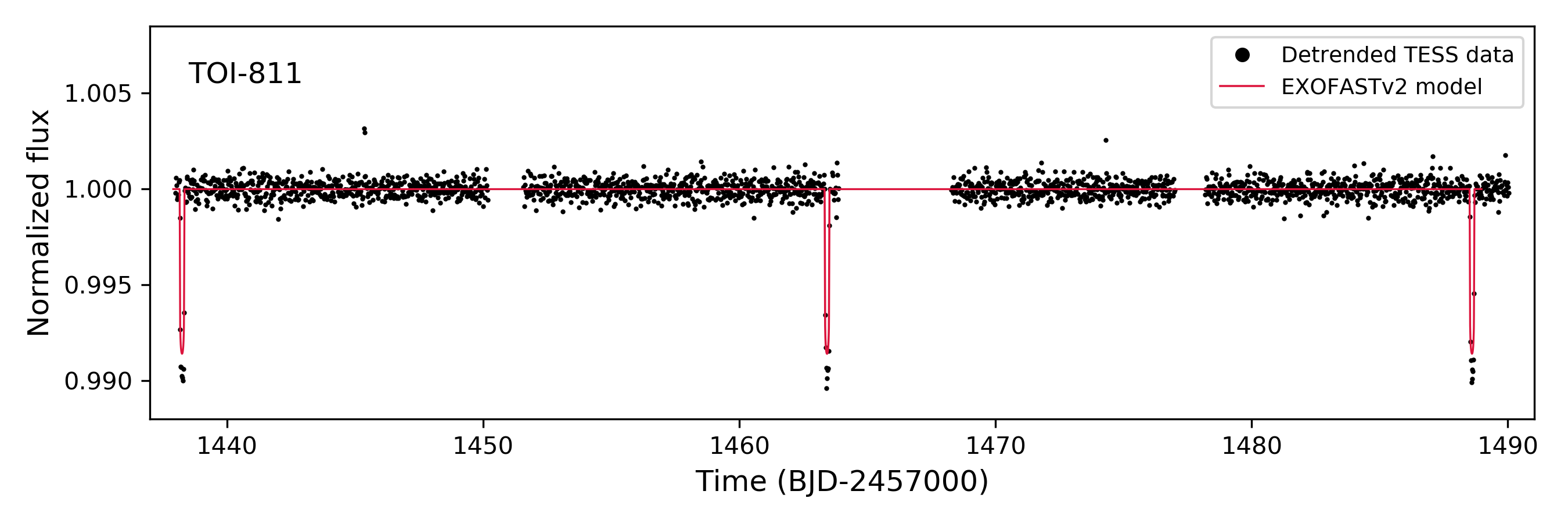}
\includegraphics[width=0.49\textwidth, trim={1.0cm 0.0cm 0.0cm 0.0cm}]{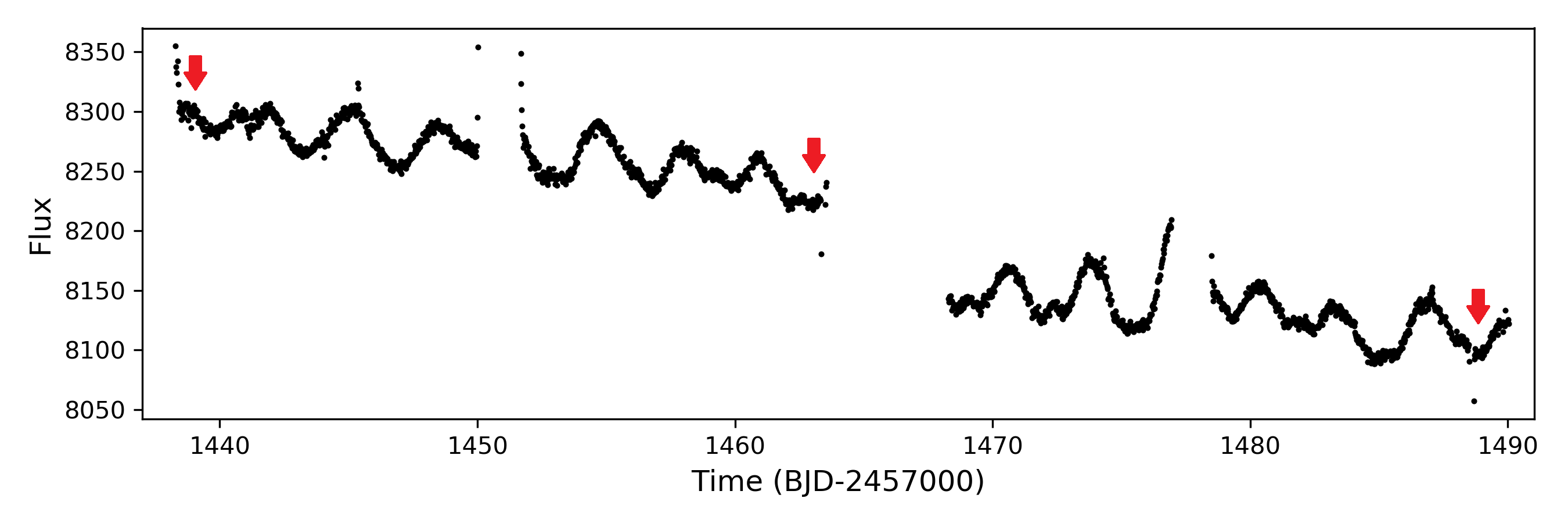}
\caption{Top: Detrended TESS light curve of TOI-811. Bottom: Raw TESS light curve of TOI-811 with locations of the transits marked by arrows. The star was observed at 30 minutes cadence in TESS Sectors 5 and 6. This star also exhibits periodic flux variations on the order of a few percent, likely due to star spots based on the changes in the patterns of the modulation; these effects have been removed for the transit analysis described in Section \ref{sec:tess_lc}.}\label{fig:lc_detrend_811}
\end{figure}

\begin{figure}[!ht]
\centering
\includegraphics[width=0.49\textwidth, trim={1.0cm 0.0cm 0.0cm 0.0cm}]{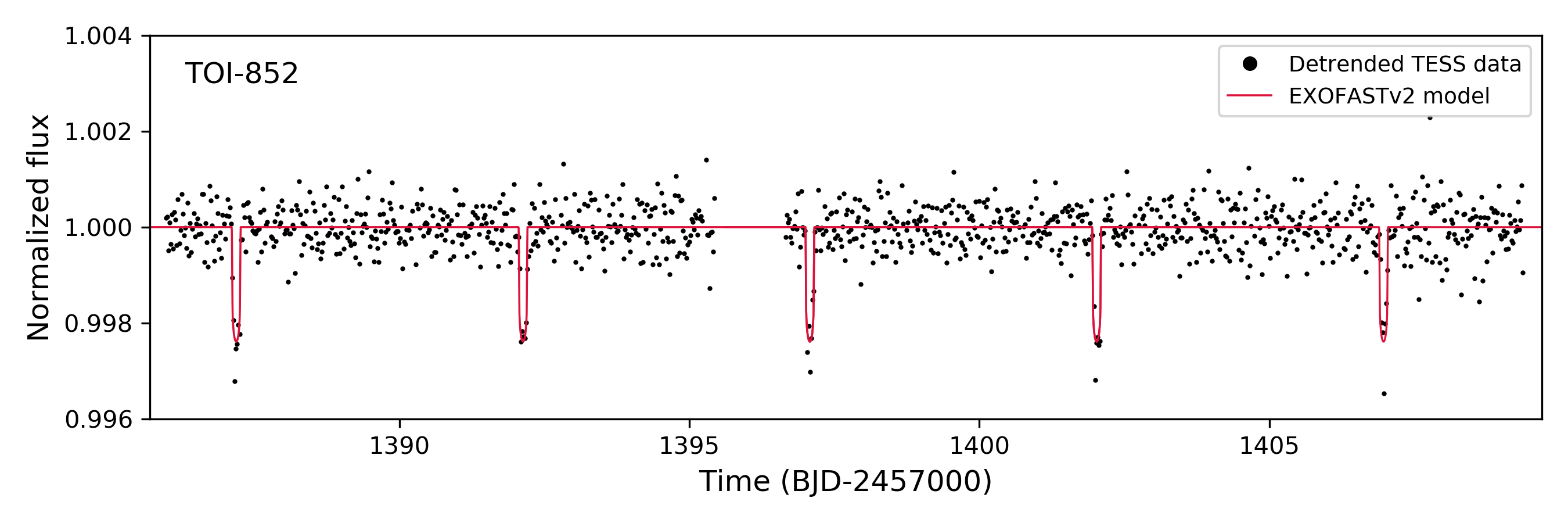}
\includegraphics[width=0.49\textwidth, trim={1.0cm 0.0cm 0.0cm 0.0cm}]{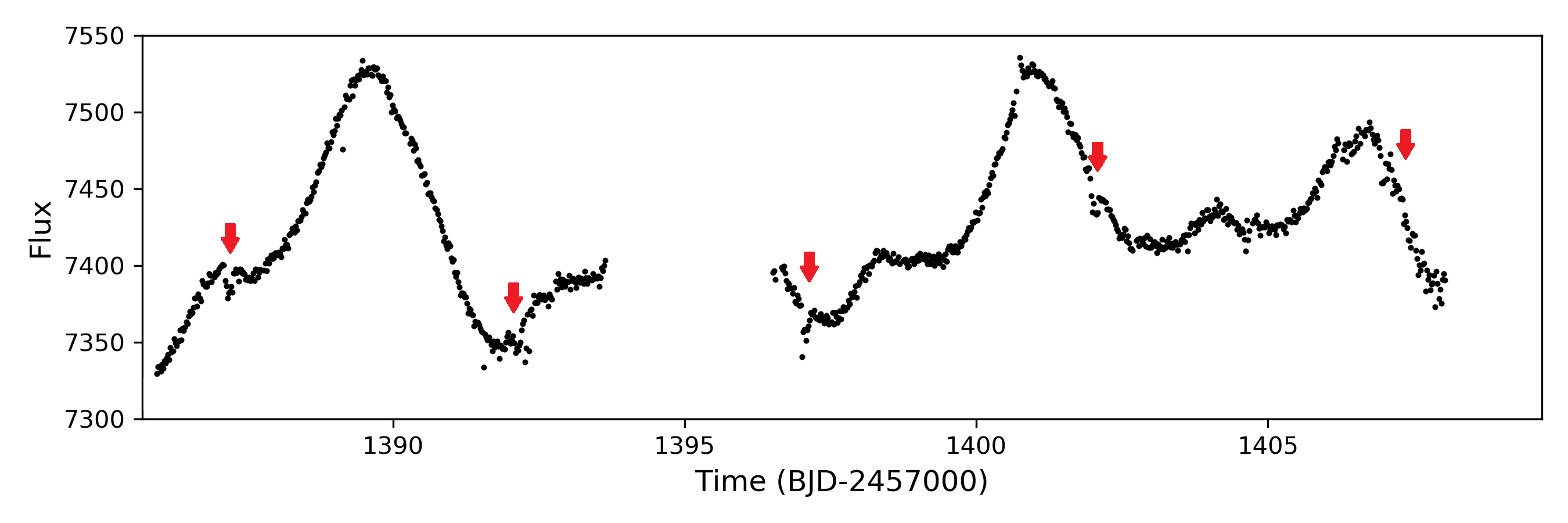}
\caption{Top: Detrended TESS light curve of TOI-852. Bottom: Raw TESS light curve of TOI-852 with locations of the transits marked by arrows. The star was observed at 30 minutes cadence in TESS Sector 3. This star also exhibits periodic 1-3\% flux variations likely due to star spots based on the changes in the patterns of the modulation; these effects have been removed for the transit analysis described in Section \ref{sec:tess_lc}.}\label{fig:lc_detrend_852}
\end{figure}

To extract the TESS light curves for analysis for each star, we use the methods described in detail in previous works using TESS data \citep[e.g.][]{vanderburg_2019, huang_2020}. The publicly available full-frame image (FFI) pixel files are calibrated by the Science Processing Operation Center \citep[SPOC,][]{jenkins_2016} and accessed via TESSCut \citep{tesscut}. We extracted light curves from 10 pixel and 9 pixel apertures for TOI-811 and TOI-852, respectively. We omit data flagged by SPOC as having poor quality. The TESS spacecraft motions are corrected by decorrelating the light curve with the spacecraft quaternion engineering data and the background flux outside the aperture, while modeling stellar variability and long-term instrumental drifts with a basis spline. The resulting light curves are shown in Figures \ref{fig:lc_detrend_811} and \ref{fig:lc_detrend_852}. Note, for the raw light curves in these figures, we do a simple 4 pixel by 4 pixel cut using the {\tt lightkurve} \citep{tesscut} extraction tool.

\subsubsection{ULMT light curve}
We observed a nearly-full transit of TOI-852b continuously for 350 minutes using 80~s exposures on 2019 October 21 in Sloan $r'$ band from the 0.61\,m University of Louisville Manner Telescope (ULMT). ULMT is located on Mt. Lemmon near Tucson, Arizona. We used the {\tt TESS Transit Finder}, which is a customized version of the {\tt Tapir} software package \citep{Jensen:2013}, to schedule our transit observations. The $4096\times4096$ SBIG STX-16803 camera has an image scale of 0.39$\arcsec$ per pixel, resulting in a $26.8\arcmin\times26.8\arcmin$ field of view. The images were defocused and have typical stellar point-spread-functions with a full-width-half-maximum (FWHM) of $4\farcs1$. The images were calibrated and photometric data were extracted with {\tt AstroImageJ} \citep{Collins:2017} using a circular aperture with radius $6\farcs6$. The light curve was detrended against airmass and FWHM. \dbf{The 1700 ppm transit was detected, and found to occur 154 minutes later than the public TOI ephemeris.} 

\subsubsection{LCOGT 1-m light curve}

We observed a partial ingress and most of the in-transit window of TOI-811b continuously for 240 minutes using 50~s exposures on 2020 March 8 in Pan-STARSS $z$-short band from the LCOGT \citep{Brown:2013} 1.0\,m node at Cerro Tololo Inter-American Observatory. We used the {\tt TESS Transit Finder} to schedule our transit observations. The $4096\times4096$ LCOGT SINISTRO cameras have an image scale of 0.389$\arcsec$ per pixel, resulting in a $26.5\arcmin\times26.5\arcmin$ field of view. The images were defocused and have typical stellar point-spread-functions with a FWHM of $6\farcs6$. The images were calibrated by the standard LCOGT {\tt BANZAI} pipeline, and photometric data were extracted with {\tt AstroImageJ} using a circular aperture with radius $8\farcs2$. The light curve was detrended against FWHM. The observation did not include baseline out-of-transit coverage, but the transit is detected on target and has a depth of at least 8000 ppm. This shows that the LCOGT observations are not inconsistent with those from TESS and MEarth. We do not include the LCOGT observations in our light curve analysis due to the lack of sufficient out of transit coverage.

\subsubsection{MEarth light curve}
The transit of TOI-811b was observed using 7 telescopes of the MEarth-South array at Cerro Tololo Inter-American Observatory (CTIO), Chile on 2020 March 8. This instrument is described in detail by \citet{2015csss...18..767I} and data reduction procedures by \citet{2007MNRAS.375.1449I,2012AJ....144..145B}.

Exposure times were 26s with all telescopes operated in focus, gathering observations continuously from the end of evening twilight until the target set below an airmass 3. Due to strong winds the resulting image quality was poor and highly variable, requiring special attention during photometric extraction and analysis.

After confirming the transit signal originated from the target star, the photometry was re-extracted with a photometric aperture radius of $r = 10$ pixels ($8\arcsec.4$ on sky) to obtain a cleaner transit curve for analysis. This large aperture does not resolve the target star from the neighboring source TIC 100757804 (2MASS J05520724-3255263) so a correction for the resulting dilution was included during analysis (details in Section \ref{sec:analysis}), based on the measured magnitude difference of $\Delta {\rm MEarth} = 1.568 \pm 0.014$ mag taking the mean and error in the mean measured from the master images (selected to have the best possible image quality) over all the telescopes. The large variation in airmass during the observation results in some residual color-dependent extinction effects in the light curve so this fit also used decorrelation against airmass.

\subsubsection{El Sauce light curve}
We observed an ingress of the transit of TOI-811 from El Sauce Observatory in Coquimbo Province, Chile. The 0.36 m telescope is equipped with a $1536\times1024$ SBIG STT-1603-3 camera with an image scale of 1$\farcs$47 per pixel resulting in a $18.8\arcmin\times12.5\arcmin$ field of view. The photometric data were extracted using {\tt AIJ}. The El Sauce data are not high enough quality to be used in our transit model analysis and El Sauce only produced in-transit data after ingress and before egress.

\begin{deluxetable}{lcccc}
 \tabletypesize{\footnotesize}
 \tablewidth{0pt}
 \tablecaption{Coordinates and magnitudes for TOI-811 and TOI-852. The $B_T$, $V_T$, $J$, $H$, $K$, WISE1, WISE2, and WISE3 values here are used to model the spectral energy distributions and constrain $T_{\rm eff}$ (along with the spectroscopic $T_{\rm eff}$ measurement) for each star. The WISE4 bandpass was not used due to a bad photometry flag. Only the TESS $T$ bandpass experiences dilution for TOI-811 and we show the non-diluted value here. \label{tab:toi_obs}}
 \tablehead{
 \colhead{} & \colhead{Description} & \colhead{TOI-811} & \colhead{TOI-852} & \colhead{Source}}
 \startdata
  & TIC IDs & TIC 100757807 & TIC 29918916 & 1\\
 $\alpha_{\rm J2000}$ &Equatorial& 05 52 07.28 & 01 38 57.66 & 1\\
 $\delta_{\rm J2000}$ &coordinates& -32 55 30.27 & -07 16 51.52 & 1\\
 $T$\dotfill & TESS $T$\dotfill & $11.123 \pm 0.006$ & $10.950 \pm 0.007$ & 2\\
 $G$\dotfill & Gaia $G$\dotfill & $11.339 \pm 0.001$ & $11.423 \pm 0.002$ & 1\\
 $B_T$\dotfill & Tycho $B_T$\dotfill & $12.098 \pm 0.203$ & $13.029 \pm 0.306$ & 3\\
 $V_T$\dotfill & Tycho $V_T$\dotfill & $11.411 \pm 0.014$ & $11.759 \pm 0.163$ &  3\\
 $J$\dotfill & 2MASS $J$\dotfill & $10.397 \pm 0.040$ & $10.278 \pm 0.022$ & 4\\
 $H$\dotfill & 2MASS $H$\dotfill & $11.147 \pm 0.050$ & $9.966 \pm 0.022$ & 4\\
 $K_S$\dotfill & 2MASS $K_S$\dotfill & $10.076 \pm 0.080$ & $9.880 \pm 0.020$ & 4\\
 WISE1\dotfill & WISE 3.4$\rm \mu m$\dotfill &  $9.462 \pm 0.060$ & $9.877 \pm 0.023$ & 5\\
 WISE2\dotfill & WISE 4.6$\rm \mu m$\dotfill & $9.471 \pm 0.060$ & $9.915 \pm 0.020$ & 5\\
 WISE3\dotfill & WISE 12$\rm \mu m$\dotfill & $9.569 \pm 0.070$ & $9.846 \pm 0.046$ & 5\\
 \hline
 \enddata
 \tablecomments{References: 1 - \cite{Lindegren2018}, 2 - \cite{stassun18}, 3 - \cite{tycho2}, 4 - \cite{2MASS}, 5 - \cite{WISE}}
\end{deluxetable}

\begin{figure}[!ht]
\centering
\includegraphics[width=0.48\textwidth, trim={0.0cm 0.0cm 0.0cm 0.0cm}]{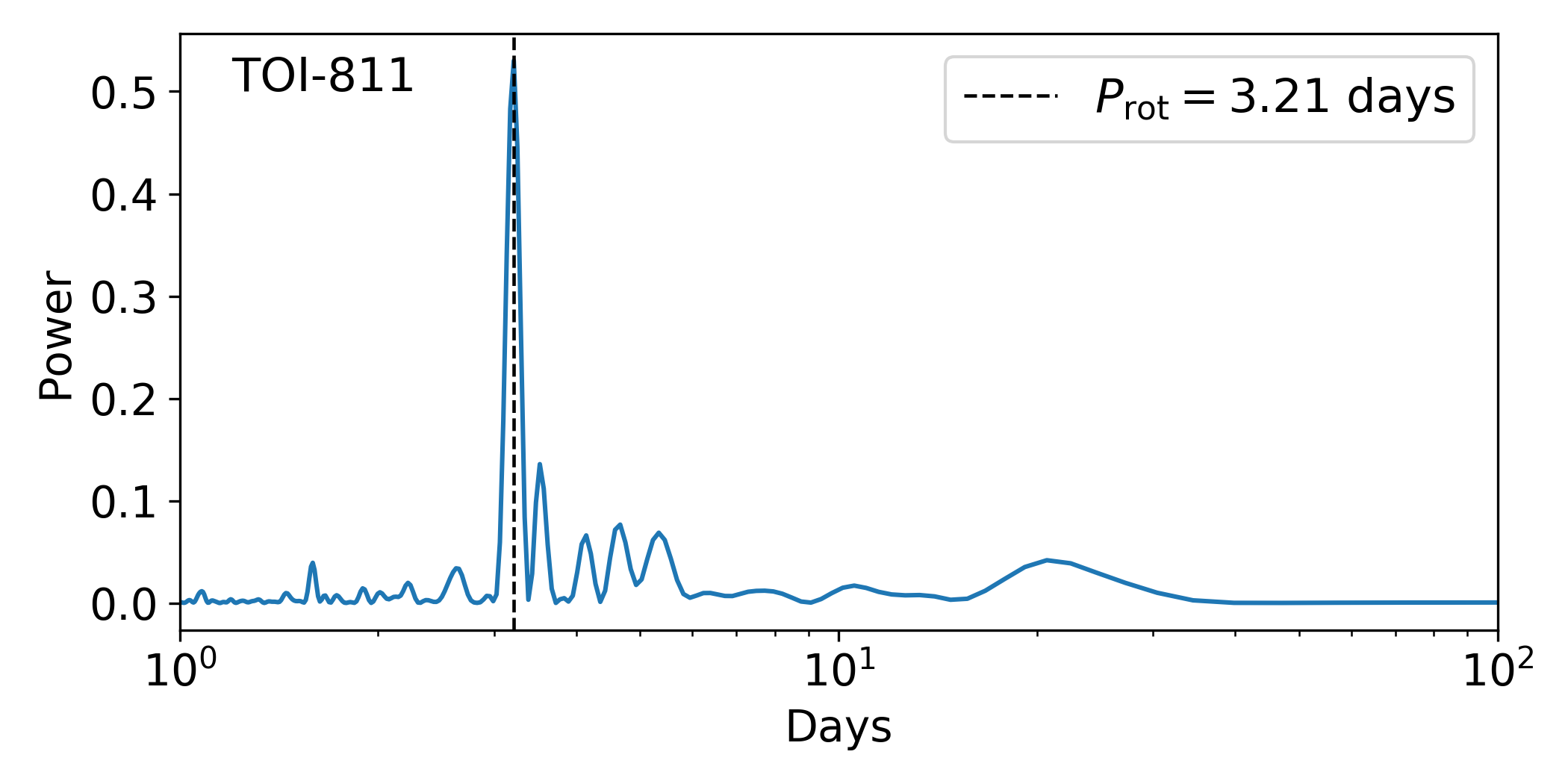}
\includegraphics[width=0.48\textwidth, trim={0.0cm 0.0cm 0.0cm 0.0cm}]{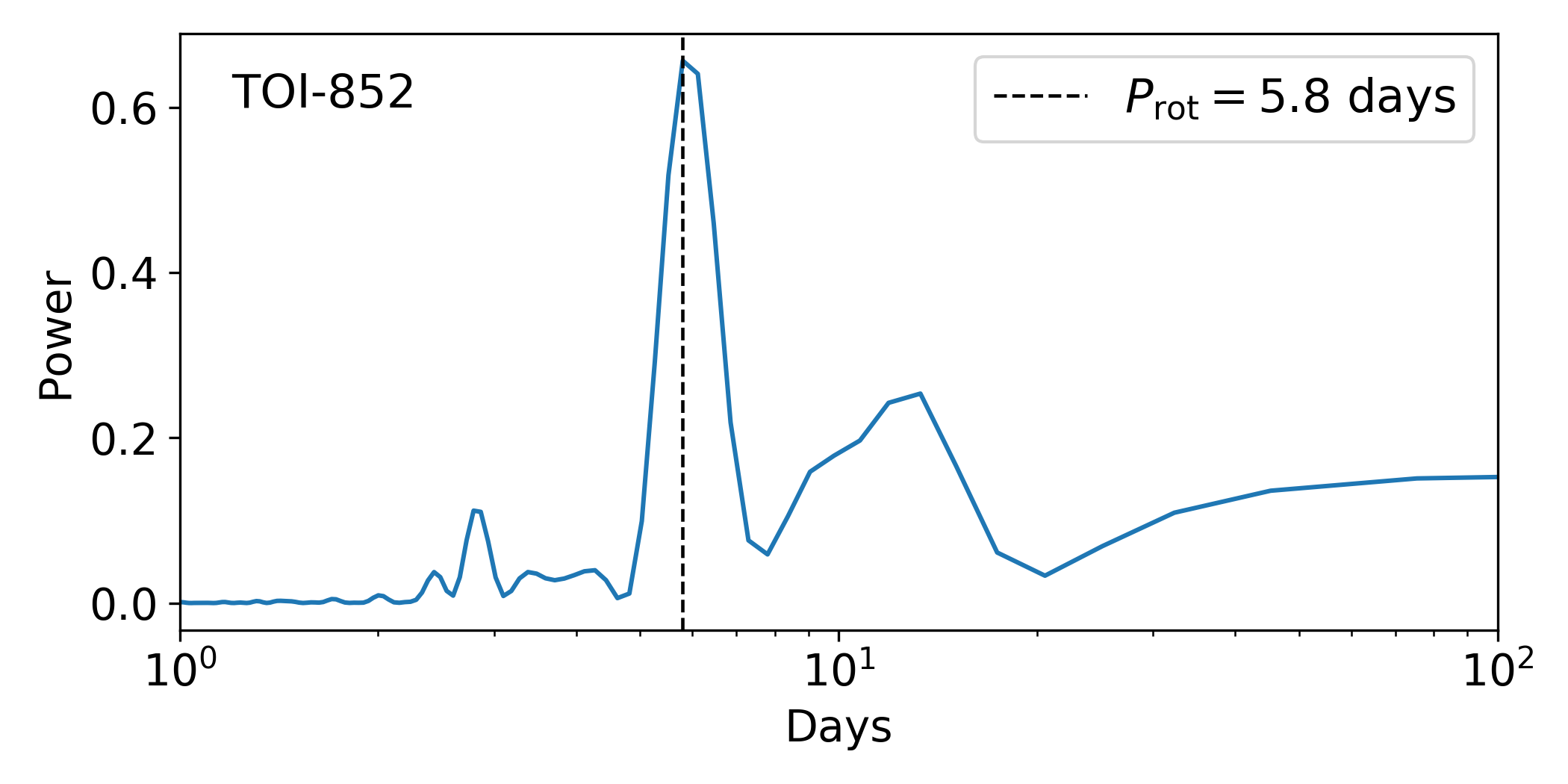}
\caption{Lomb-Scargle periodograms from the TESS light curve of TOI-811 (top) and TOI-852 (bottom). The TESS periodogram indicates a peak frequency at $3.21\pm 0.02$ days for TOI-811 and a peak frequency at $5.80\pm 1.39$ days for TOI-852 in the 1-100 day range. The uncertainties on the rotation period are determined from the FWHM of the peak from the Lomb-Scargle analysis.}\label{fig:periodogram}
\end{figure}

\begin{figure}[!ht]
\centering
\includegraphics[width=0.49\textwidth, trim={0.0cm 0.0cm 0.0cm 0.0cm}]{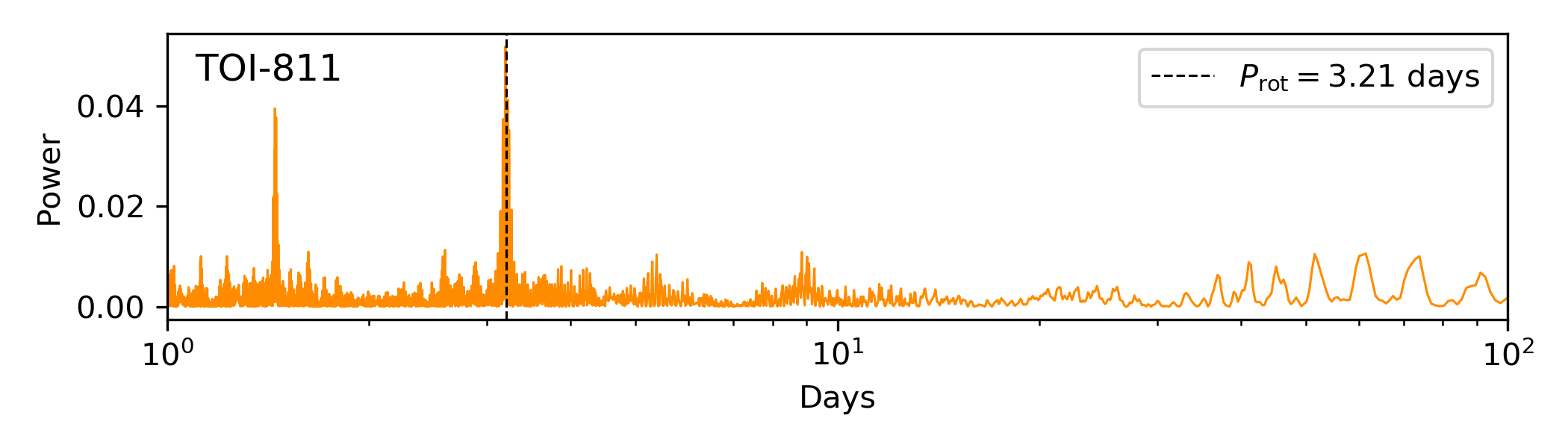}
\includegraphics[width=0.49\textwidth, trim={0.0cm 0.0cm 0.0cm 0.0cm}]{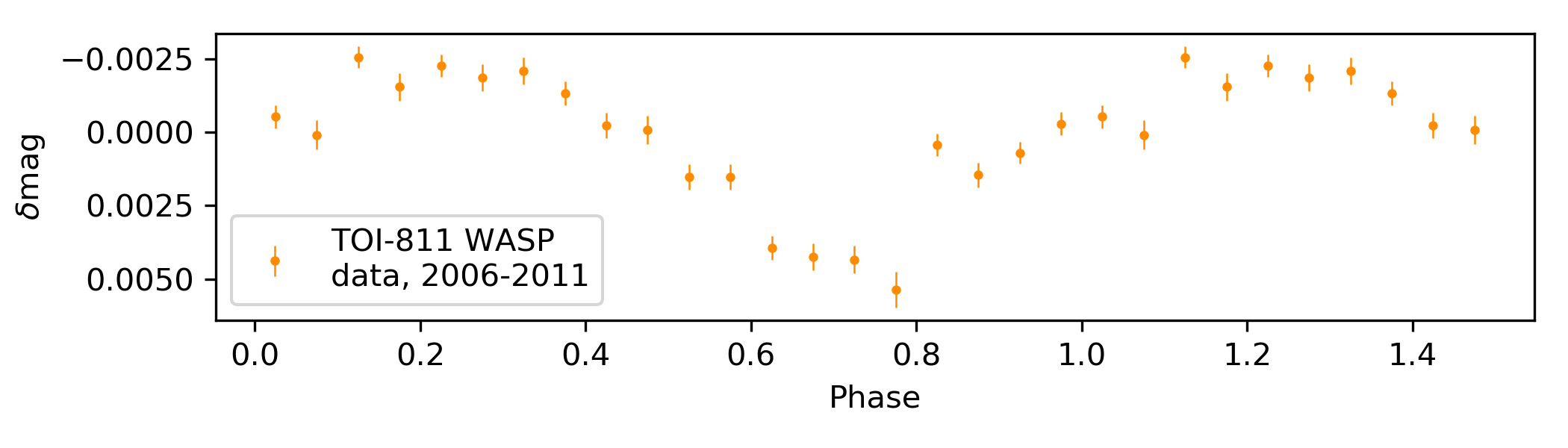}
\includegraphics[width=0.49\textwidth, trim={0.0cm 0.0cm 0.0cm 0.0cm}]{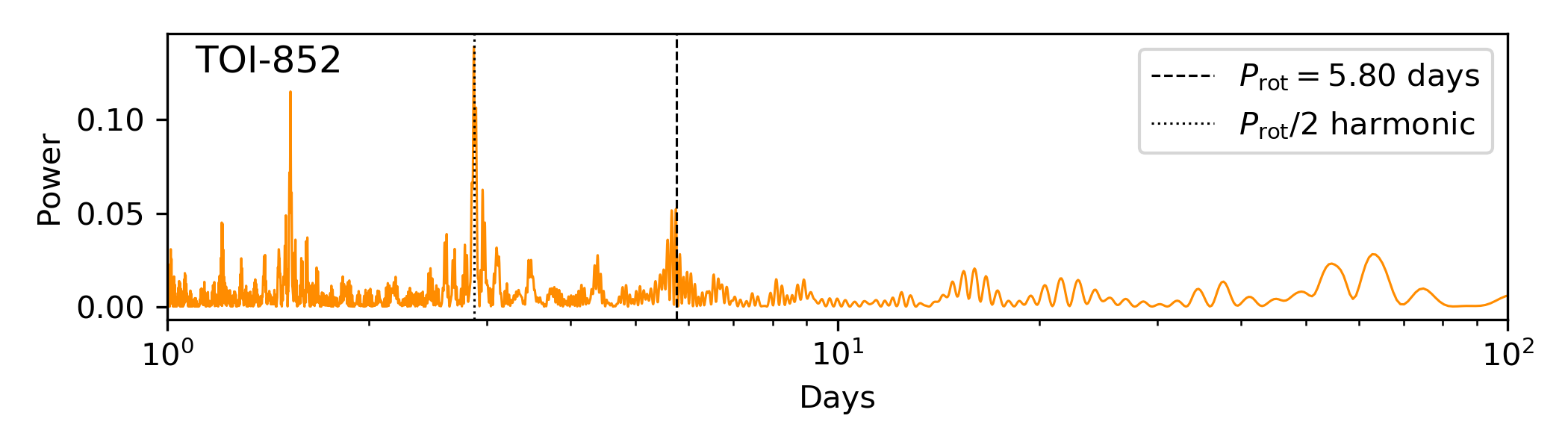}
\includegraphics[width=0.49\textwidth, trim={0.0cm 0.0cm 0.0cm 0.0cm}]{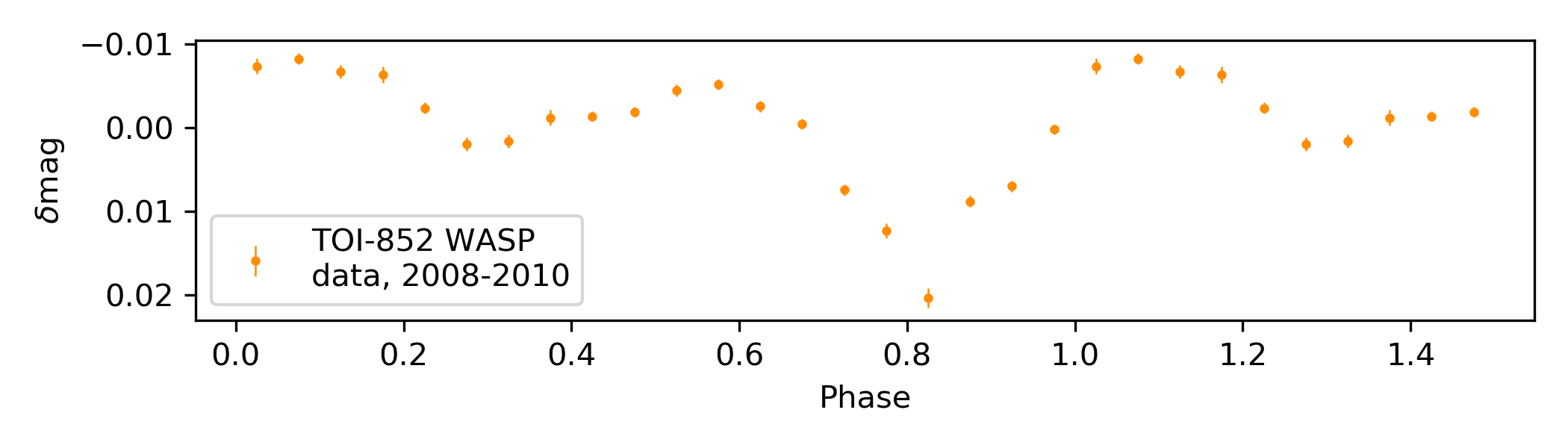}
\caption{Phase folded WASP light curves and corresponding Lomb-Scargle periodograms for TOI-811 (top two panels) and TOI-852 (bottom two panels). The WASP periodogram for TOI-852 reveals a harmonic of the stellar rotation period of $P_{\rm rot}=5.80 \pm 1.39$ days that the TESS data confirm. Only with the addition of the TESS data was the smaller peak at 5.80 days chosen. The phased data show 1.5 phases to make the variation in the WASP data easier to see. The WASP light curve on TOI-811 show a peak-to-peak separation of approximately one period while the light curve for TOI-852 shows some substructure between one period (possibly from star spots on the back face of the star), which is reflected in the harmonic in the Lomb-Scargle periodogram.}\label{fig:wasp}
\end{figure}

\subsubsection{Photometric modulation \& stellar rotation period}
Before the TESS mission detected the transits of the BDs in the TOI-811 and TOI-852 systems, the Wide Angle Search for Planets (WASP) survey found photometric modulation in the light curves of both stars. The period of the modulation calculated from the WASP light curve of each star is consistent with the period calculated from the raw TESS light curves, which are shown in Figures \ref{fig:lc_detrend_811} and \ref{fig:lc_detrend_852}. We interpret the cause of the photometric modulation to be a result of star spots coming into and out of view with the rotation of each star. Thus, we may use this variation in each star's brightness to derive a rotation period for the stars. Using the TESS data, for TOI-811, we find a rotation period of $P_{\rm rot}=3.21 \pm 0.02$ days, and for TOI-852, we find a $P_{\rm rot}=5.80 \pm 1.39$ days. The results of the Lomb-Scargle periodogram analysis we performed on each star's TESS light curve to derive these periods are shown in Figure \ref{fig:periodogram}.

One benefit to the WASP data is that they span a much longer time frame than the TESS sectors do for each star. The WASP observations of TOI-811 span from 2006 to 2011 and those for TOI-852 span from 2008-2010. Though the WASP data reveal a consistency in the measured rotation period when compared to the TESS data, the WASP light curves do not have the sensitivity to detect the transits of either TOI-811b or TOI-852b, so the TESS data are still very important in this regard. We perform a Lomb-Scargle analysis on the WASP data and fold the data at the dominant peak period (3.21 days) for TOI-811 and the mutual strong candidate peak period (5.80 days for the TESS and WASP data) for TOI-852. The phase folded WASP light curves at these periods are shown in Figure \ref{fig:wasp}. Together, the WASP and TESS periodograms help to rule out any harmonics that appear in only one data set for TOI-852.

\subsection{High resolution imaging and contaminating sources}

Though the ground-based photometric follow up of both stars allows us to confirm that the transits originate from the target stars, we also pursue observations at resolutions that are higher than seeing-limited in order to see if there are any very nearby companions and determine whether or not they contaminate the photometry of the target star. This is where we make use of speckle imaging with the 4.1-m Southern Astrophysical Research (SOAR) telescope \citep{tokovinin2018_soar} located at the Cerro Tololo Inter-American Observatory. We use SOAR speckle imaging to confirm whether or not there are other objects of comparable brightness within a $3\farcs0$ field of view, which is much smaller than that of the aperture size used by the TESS mission. Such bright objects may significantly influence the RVs we measure and the broadband photometry of the target star, which may produce false positives or otherwise distort the mass and radius determinations for the companions. In the case of our search for transiting BDs, we are cautious against contaminating stars that dilute the depth of a transit by enough to alter the measured radius of the BD.

\begin{deluxetable*}{cccccccc}
\tabletypesize{\footnotesize}
\tablewidth{0pt}

 \tablecaption{Nearby sources to TOI-811 from Gaia DR2 data. This table lists sources within $30\arcsec$ of TOI-811 that are brighter than $G=19$. The parallaxes ($\varpi$), proper motions ($\mu_{\rm \alpha}$, $\mu_{\rm \delta}$), and \dbf{systemic radial velocities ($\gamma$) of TIC 100757804 and TOI-811 indicate that they are comoving.} Using the most up-to-date coordinates from the TESS Input Catalog v8 \citep{TICv8}, we determine the angular separation between the primary and secondary stars in TOI-811 to be $4\farcs35$. \label{tab:gaia}}

 \tablehead{
 \colhead{TIC ID} & \colhead{$\alpha$ (J2015.5)} & \colhead{$\delta$ (J2015.5)}  & \colhead{$\varpi$ (mas)} & \colhead{$\mu_{\rm \alpha}$ ($\rm mas\,yr^{-1}$)} & \colhead{$\mu_{\rm \delta}$ ($\rm mas\, yr^{-1}$)} & $G$ (mag) & $\gamma$ ($\rm kms^{-1}$)}

\startdata 
100757807 & 05 52 07.28 & -32 55 30.27
 & $3.4923 \pm 0.0211$ & $14.968 \pm 0.036$ & $25.624 \pm 0.045$ & 11.34 & $34.2 \pm 0.2$\\
100757804 & 05 52 07.26 & -32 55 25.94 & $3.4943 \pm 0.0144$ & $15.897 \pm 0.023$ & $24.405 \pm 0.027$ & 13.17 & $37.0 \pm 3.0$\\
100757805 & 05 52 08.68 & -32 55 27.50 & $0.7957 \pm 0.0357$ & $-0.369 \pm 0.066$ & $0.664 \pm 0.070$ & 16.11 & -\\
\enddata
\end{deluxetable*}

\begin{figure}[!ht]
\centering
\includegraphics[width=0.45\textwidth, trim={0.0cm 0.0cm 0.0cm 0.0cm}]{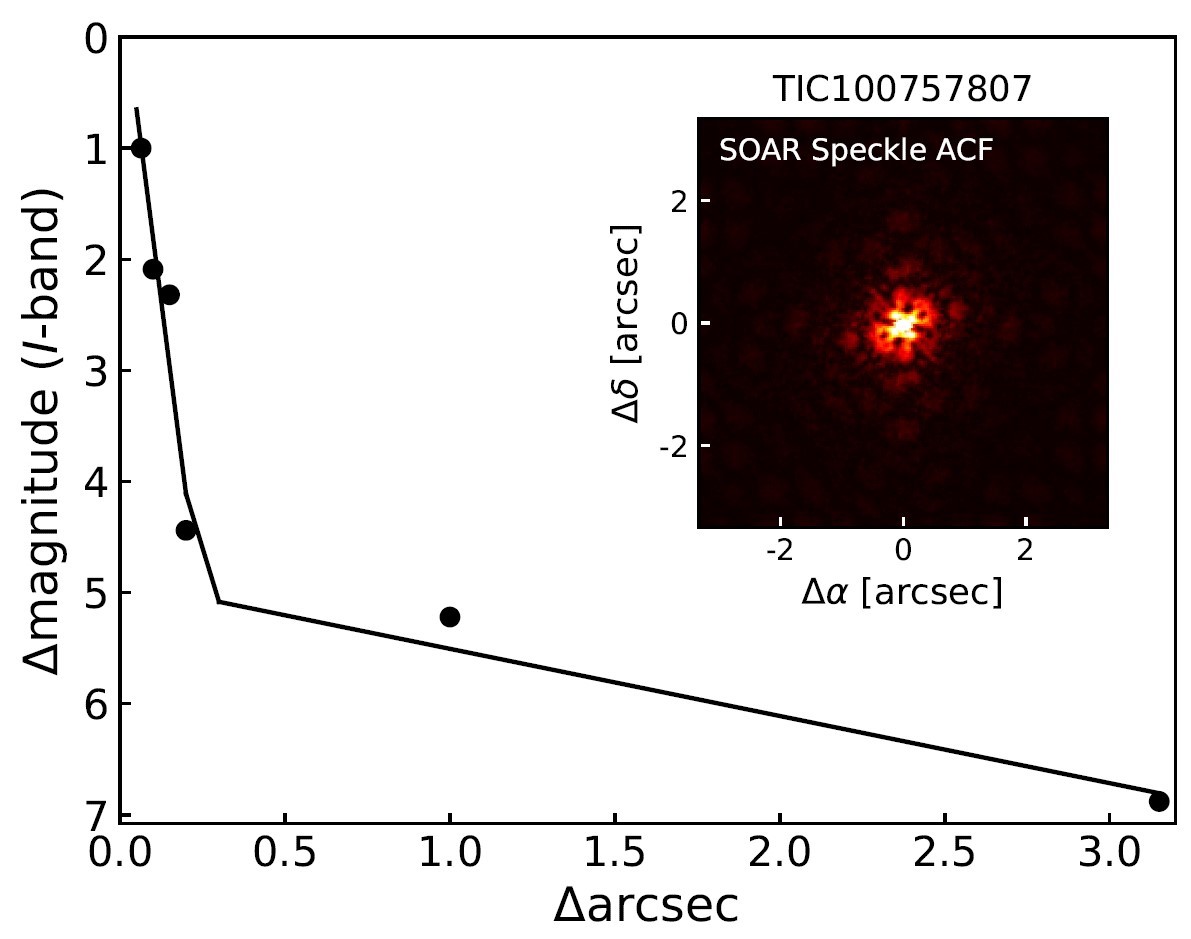}
\includegraphics[width=0.45\textwidth, trim={0.0cm 0.0cm 0.0cm 0.0cm}]{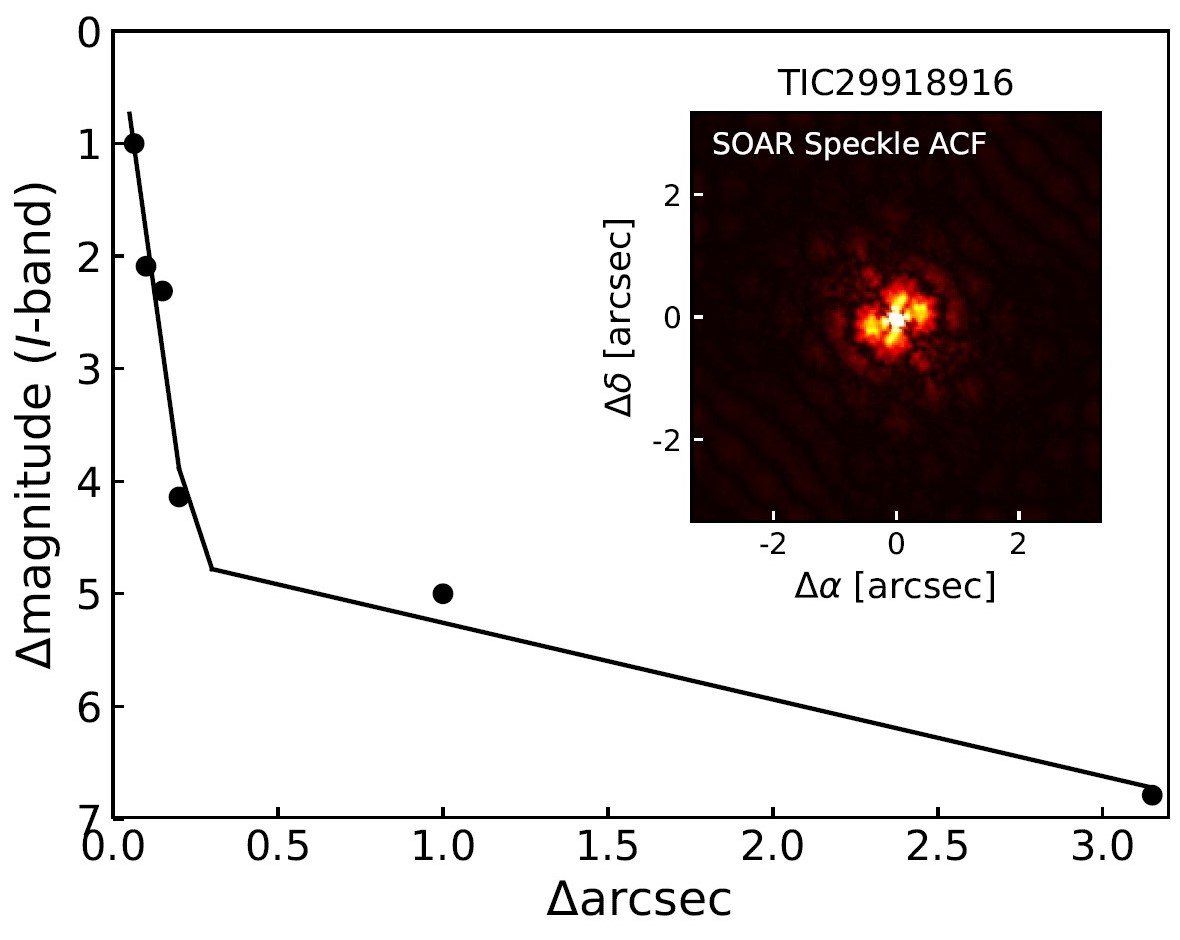}
\caption{The 5$\sigma$ sensitivity limits and auto-correlation functions of the SOAR speckle observations of TOI-811 (top) and TOI-852 (bottom). The black circles are measured data points and the lines are fits in two different separation regimes. No nearby contaminating sources are detected within 3$\arcsec$, but data from Gaia DR2 confirm a companion source to TOI-811 at roughly $4\farcs35$.}\label{fig:soar}
\end{figure}

On 2019 August 12, we took SOAR speckle observations of both TOI-811 and TOI-852 in the Cousin-\textit{I}-band, which is a similar bandpass to that of TESS. Further details of how SOAR observations are carried out are available in \cite{ziegler2019_soar}. The 5$\sigma$ detection sensitivity and speckle auto-correlation functions from the observations are shown in Figure \ref{fig:soar}. No nearby stars were detected within 3$\arcsec$ of either TOI-811 or TOI-852 in the SOAR observations. However, when using data from Gaia DR2, we see a companion star to TOI-811 at roughly $4\farcs35$ away. This nearby star is only about 5 times fainter than the primary star, so we carefully take contamination effects into account in our analysis discussed in later sections. For a general sense of the relative positions and magnitudes of the primary and secondary stars in the TOI-811 system, see Table \ref{tab:gaia}. We find no bright sources beyond 3$\arcsec$ and out to 30$\arcsec$ for TOI-852.

\subsection{TRES spectra}\label{sec:tres}
We use the TRES instrument on Mt. Hopkins, Arizona to obtain spectra for both TOI-811 and TOI-852. TRES has a resolving power of $R\sim 44\,000$ and covers a wavelength range of $\rm 3900\AA$ to $\rm 9100\AA$. We use multiple echelle orders for each spectrum to measure a relative RV at each phase in the orbit of the transiting BD. We visually review each order to omit those with low signal-to-noise per resolution element (S/N) and to remove cosmic rays. Each order is cross-correlated with the highest observed S/N spectrum of the target star and then the average RV of all the orders per spectrum is taken as the RV of the star for that observation.

We use the stellar parameter classification (SPC) software package by \cite{spc} to derive $T_{\rm eff}$, metallicity, $\log{g}$, and the projected stellar equatorial velocity $v\sin{I_\star}$ from co-added TRES spectra of TOI-811 and TOI-852. SPC uses a library of calculated spectra in the $\rm 5030-5320\AA$ wavelength range, centered near the Mg b triplet.

For TOI-811, we took two spectra to confirm that the transiting candidate BD is indeed in the BD mass range. These spectra were taken on 2019 October 27 and 2019 November 9 at roughly opposite quadrature with exposure times of 3000\,s and 2500\,s, respectively, giving us a S/N between 27 and 36. Using SPC, we derive the following stellar parameters for TOI-811: $T_{\rm eff} = 6013 \pm 56$K, $\log{g} = 4.31 \pm 0.10$, $\rm [Fe/H] = +0.41 \pm 0.08$, and $v\sin{I_\star}= 7.11 \pm 0.5$ $\rm km\, s^{-1}$.

For TOI-852, we took a series of 11 spectra to derive an orbital solution for the system. The exposure times for these follow up spectra range from 1600\,s to 3600\,s to give a S/N range of 21 to 36. The stellar parameters for these spectra derived with SPC for TOI-852 are: $T_{\rm eff} = 5746 \pm 51$K, $\log{g} = 4.29 \pm 0.10$, $\rm [Fe/H] = +0.30 \pm 0.08$, and $v\sin{I_\star}= 14.5 \pm 0.5$ $\rm km\, s^{-1}$. For consistency, we use the $T_{\rm eff}$ and [Fe/H] values only from SPC for TOI-811 and TOI-852 to set our priors for the global analysis discussed in the next section.

\subsection{CHIRON spectra}\label{sec:chiron}
To characterize the RVs of TOI-811, we obtained a series of 11 spectroscopic observations using the CHIRON spectrograph on the 1.5\,m SMARTS telescope \citep{chiron}, located at Cerro Tololo Inter-American Observatory, Chile. CHIRON is a high resolution echelle spectrograph that is fed via an image slicer and a fiber bundle. CHIRON has a resolving power of $\lambda / \Delta \lambda \equiv R \sim 80\,000$ over the wavelength region $\rm 4100\AA$ to $\rm 8700\AA$. The wavelength calibration is obtained via thorium-argon hollow-cathode lamp exposures that bracket each stellar spectrum.

To derive the stellar RVs, we performed a least-squares deconvolution \citep{1997MNRAS.291..658D} between the observed spectra and a non-rotating synthetic template generated via ATLAS9 atmospheric models \citep{Castelli:2004} at the stellar atmospheric parameters of each target. We then model the stellar line profiles derived from the least-squares deconvolution via an analytic rotational broadening kernel as per \cite{2005oasp.book.....G}. The derived RVs for TOI-811 are listed in Table~\ref{tab:rvs}.

\subsection{CORALIE spectra}\label{sec:coralie}
We obtained 11 spectroscopic observations of TOI-852 from 2019 August 4 to 2019 August 26 with the high resolution CORALIE spectrograph on the Swiss 1.2 m Euler telescope at the La Silla Observatory, Chile \citep{Queloz2001}. CORALIE has a resolving power of $R\sim 60\,000$ and is fed by two fibers: a 2$\arcsec$ on-sky science fiber encompassing the star and another fiber that can either connect to a Fabry-P\'erot etalon for simultaneous wavelength calibration or on-sky for background subtraction of sky flux. We observed TOI-852 in the simultaneous Fabry-P\'erot wavelength calibration mode using exposure times of 1800\,s. The spectra were reduced with the CORALIE standard reduction pipeline and RVs were computed for each epoch by cross-correlating with a binary G2 mask \citep{Pepe2002}. Bisector-span, full width at half-maximum (FWHM), and other cross-correlation function (CCF) line-profile diagnostics were computed as well using the standard CORALIE pipeline. The projected rotational velocity of the star, $v\sin{I_\star}$, for TOI-852 was computed using the calibration between $v\sin{I_\star}$ and the width of the CORALIE CCF from \cite{Santos2002}. 

We derived stellar parameters including effective temperature $T_{\rm eff}$, surface gravity $\log{g}$, and metallicity $\rm [Fe/H]$ using SpecMatch-Emp \citep{Yee2017} on stacked CORALIE spectra of TOI-852. SpecMatch-Emp uses a library of stars with well-determined parameters to match the input spectra and derive spectral parameters. We use a spectral region that includes the Mg I b triplet ($\rm 5100\AA-5400\AA$) to match our spectra. SpecMatch-Emp uses $\chi^{2}$ minimization and a weighted linear combination of the five best matching spectra in the SpecMatch-Emp library to determine $T_{\rm eff}$, $\log{g}$, and $\rm [Fe/H]$. From CORALIE, we find $T_{\rm eff} = 5646 \pm 110$K, $\log{g} = 4.17 \pm 0.12$, $\rm [Fe/H] = +0.17 \pm 0.09$, and $v\sin{I_\star}= 15.0 \pm 0.8$ $\rm km\, s^{-1}$ for TOI-852. We compare these CORALIE values to those found with SPC and the TRES spectra in Table \ref{tab:spec_values}.

\subsection{Goodman spectra of TIC 100757804}\label{sec:goodman}
\dbf{On the night of 2020 October 25, we observed TIC 100757804 (the nearby star to TOI-811) with the Goodman High-Throughput Spectrograph \citep{Goodman} on the SOAR 4.1\,m telescope atop Cerro Pachón, Chile. We took all exposures using the red camera, the 1200 l/mm grating, the M5 setup, and a $0.46\farcs$ slit rotated to the parallactic angle, which provided a resolving power of $R\sim5000$ spanning $\rm 6350-7500\AA$. We took five exposures, each with 300\,s integration times. Due to a slow drift in the wavelength solution of Goodman, we took a set of Ne arcs immediately before and after the target.} 

\dbf{Using custom scripts, we performed bias subtraction, flat fielding, optimal extraction of the target spectrum, and mapped pixels to wavelengths using a fifth-order polynomial derived from the Ne lamp spectra. We then stacked the five extracted spectra using the robust weighted mean. The stacked spectrum had a signal-to-noise ratio $>100$ over most of the wavelength range.}

\dbf{We measured the radial velocity of 2MASS J05520724-3255263 by cross-correlating the Goodman spectrum against a series of radial-velocity templates from \citet{Nidever:2002} taken with Goodman using an identical setup. This gave a velocity of $37.0\pm 3.0$km\,s$^{-1}$; with errors largely limited by Goodman's wavelength stability. We thus confirm that the systemic velocity of TIC 100757804 is consistent with that of TOI-811 at $\gamma_{\rm CHIRON} = 34.2\pm 0.2{\rm kms^{-1}}$.}

\begin{deluxetable*}{ccccc}
\tabletypesize{\footnotesize}
\tablewidth{0pt}

\tablecaption{Relative radial velocities of TOI-811 from CHIRON and of TOI-852 from TRES and CORALIE. The offsets $\gamma$ for each instrument are free parameters in our {\tt EXOFASTv2} analysis. \label{tab:rvs}}

 \tablehead{
 \colhead{$\rm BJD_{\rm TDB}-2450000$} & \colhead{RV ($\rm m\, s^{-1}$)} & \colhead{$\sigma_{\rm RV}$ ($\rm m\, s^{-1}$)} & \colhead{Instrument} & \colhead{Target}}

\startdata
8582.55031 & 36725 & 34.8 & CHIRON & TOI-811\\
8611.47375 & 35730 & 31.9 & CHIRON & TOI-811\\
8774.83500 & 34460 & 43.7 & CHIRON & TOI-811\\
8777.82844 & 35657 & 43.3 & CHIRON & TOI-811\\
8785.84417 & 36535 & 40.8 & CHIRON & TOI-811\\
8789.80353 & 32951 & 35.2 & CHIRON & TOI-811\\
8793.85537 & 30425 & 19.9 & CHIRON & TOI-811\\
8824.72234 & 34298 & 45.8 & CHIRON & TOI-811\\
8829.73299 & 36221 & 47.0 & CHIRON & TOI-811\\
8852.76285 & 35644 & 40.5 & CHIRON & TOI-811\\
8857.62889 & 36734 & 39.3 & CHIRON & TOI-811\\
8863.68695 & 35438 & 37.3 & CHIRON & TOI-811\\
8873.64771 & 33312 & 39.9 & CHIRON & TOI-811\\
\hline
8731.866192 &   4442  &    80.3 & TRES & TOI-852\\
8773.847301 &  -4827  &   168.0 & TRES & TOI-852\\
8776.896751 &   4716  &    65.1 & TRES & TOI-852\\
8777.726909 &    781  &    94.8 & TRES & TOI-852\\
8778.796702 &  -5021  &   115.8 & TRES & TOI-852\\
8779.812572 &  -3729  &    74.1 & TRES & TOI-852\\
8780.723337 &   1880  &    88.7 & TRES & TOI-852\\
8782.797983 &    -93  &   107.8 & TRES & TOI-852\\
8783.827445 &  -5330  &   117.9 & TRES & TOI-852\\
8784.727440 &  -3713  &   120.7 & TRES & TOI-852\\
8785.740227 &   2456  &   158.7 & TRES & TOI-852\\
8786.733614 &   4591  &    86.1 & TRES & TOI-852\\
8787.730941 &   0.00   &  107.8 & TRES & TOI-852\\
8699.861071 & -21937 &  70.5  & CORALIE & TOI-852\\
8702.770755 & -12053 &  70.1  & CORALIE & TOI-852\\
8704.878168 & -22141 &  102.8 & CORALIE & TOI-852\\
8708.739945 & -17495 &  103.2 & CORALIE & TOI-852\\
8712.871395 & -12913 &  156.9 & CORALIE & TOI-852\\
8713.755792 & -18147 &  95.9  & CORALIE & TOI-852\\
8714.887268 & -22067 &  65.9 & CORALIE & TOI-852\\
8717.791778 & -12588 &  107.5 & CORALIE & TOI-852\\
8718.904775 & -19142 &  76.8 & CORALIE & TOI-852\\
8720.887670 & -17739 &  82.8 & CORALIE & TOI-852\\
8721.754744 & -12666 &  69.3 & CORALIE & TOI-852\\
\enddata
\end{deluxetable*}

\begin{deluxetable}{cccc}
\tabletypesize{\footnotesize}
\tablewidth{0pt}

 \tablecaption{Spectroscopic values for TOI-852 from CORALIE and TRES compared to {\tt EXOFASTv2} results. Only the spectroscopic $T_{\rm eff}$ and [Fe/H] from TRES are used as Gaussian priors in the {\tt EXOFASTv2} analysis. \label{tab:spec_values}}

 \tablehead{
 \colhead{Parameter} & \colhead{CORALIE} & \colhead{TRES} & \colhead{{\tt EXOFASTv2}}}

\startdata 
$T_{\rm eff}$ (K) &  $5646 \pm 110$ & $5746 \pm 51$ & $5768 \pm 84$\\
$\log{g}$ (cgs) &  $4.17 \pm 0.12$ & $4.29 \pm 0.10$ & $4.09 \pm 0.04$\\
$\rm [Fe/H]$ (dex) &  $+0.17 \pm 0.05$ & $+0.30 \pm 0.08$ & $+0.33 \pm 0.09$ \\
$v\sin{I_\star}$ ($\rm km\,s^{-1}$) &  $15.0 \pm 0.8$ & $14.5 \pm 0.5$ & - \\
$R$ (resolving power) & 60,000 & 44,000 & - \\
\enddata
\end{deluxetable}

\begin{figure}[!ht]
\centering
\includegraphics[width=0.45\textwidth, trim={0.0cm 0.0cm 0.0cm 0.0cm}]{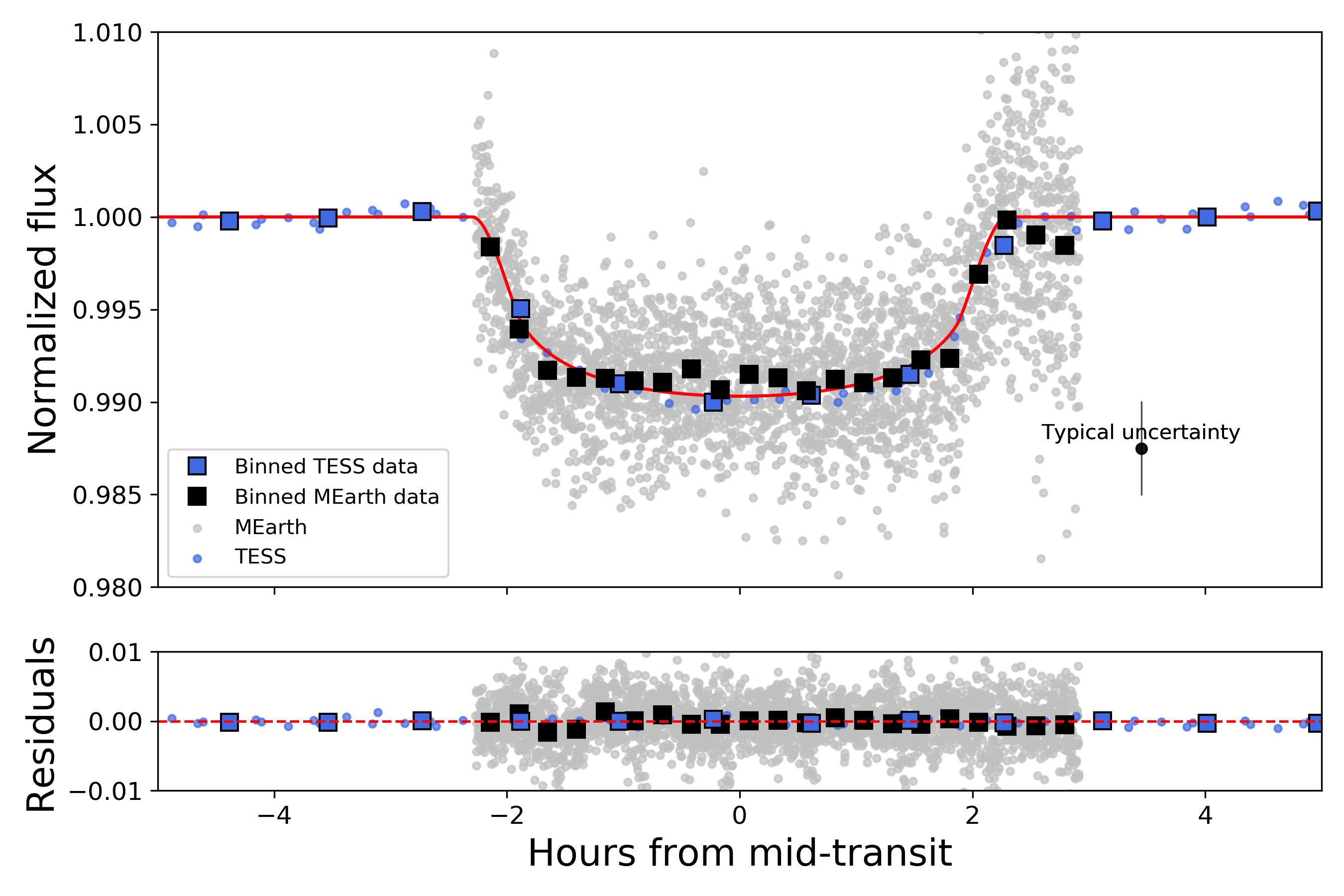}
\includegraphics[width=0.45\textwidth, trim={0.0cm 0.0cm 0.0cm 0.0cm}]{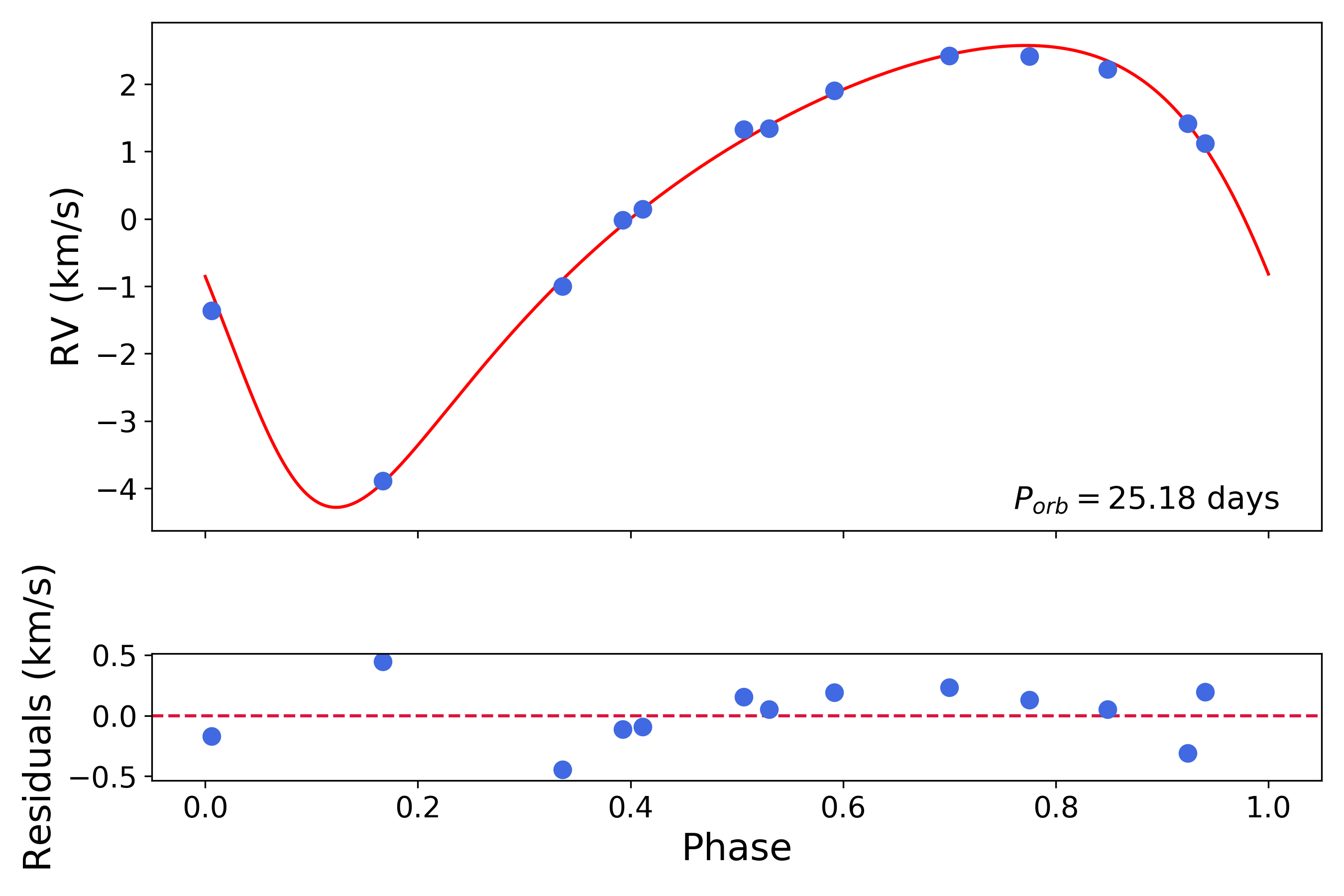}
\caption{Top: TESS and MEarth light curves of TOI-811 with global {\tt EXOFASTv2} transit model in red. The binning for the TESS data uses bins averaged over 150 minutes and the binning for the MEarth data are averaged over 80 minutes. Bottom: CHIRON multi-order relative radial velocities of TOI-811 with {\tt EXOFASTv2} orbital solution plotted in red. }\label{fig:toi811_obs}
\end{figure}

\begin{figure}[!ht]
\centering
\includegraphics[width=0.45\textwidth, trim={0.0cm 0.0cm 0.0cm 0.0cm}]{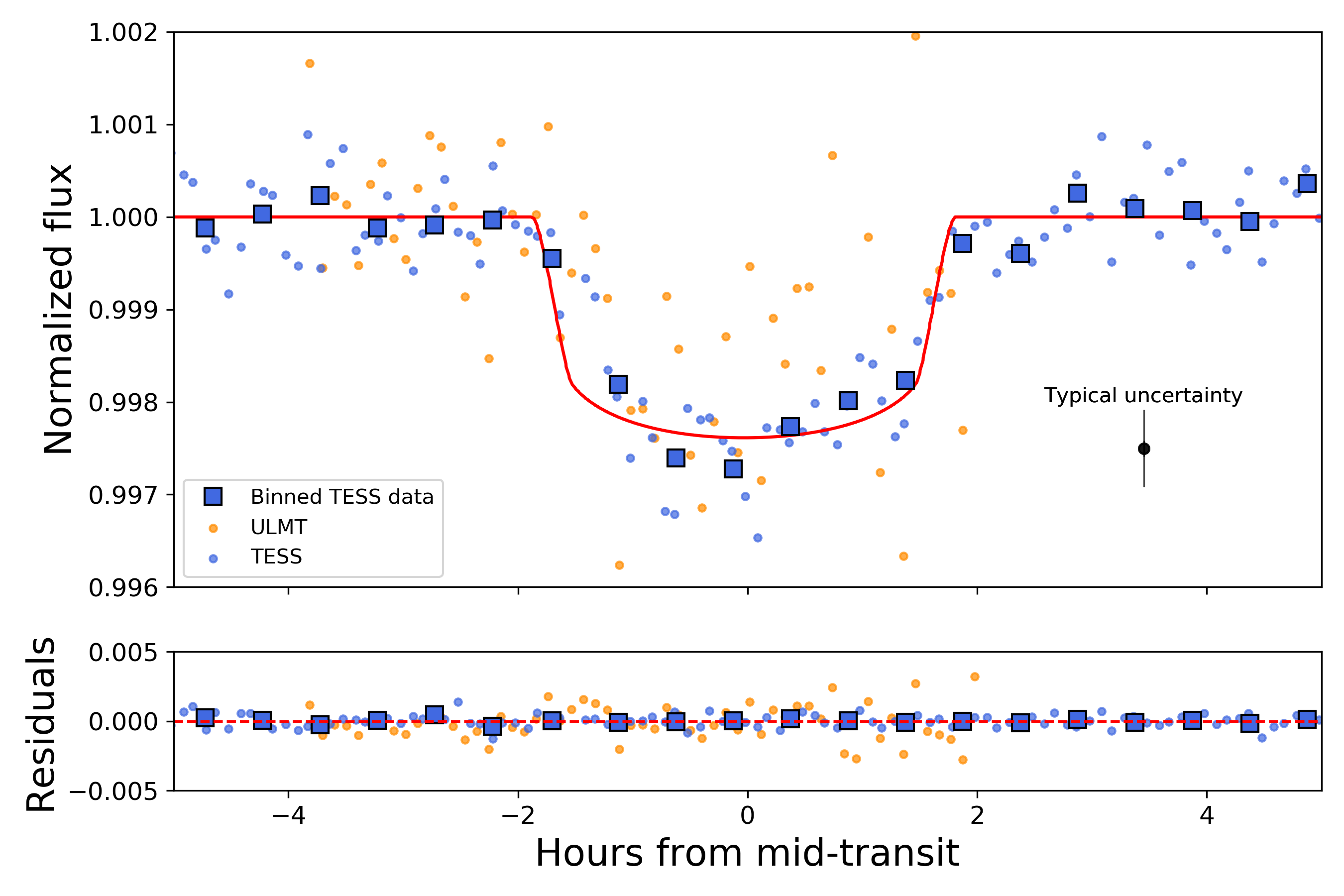}
\includegraphics[width=0.45\textwidth, trim={0.0cm 0.0cm 0.0cm 0.0cm}]{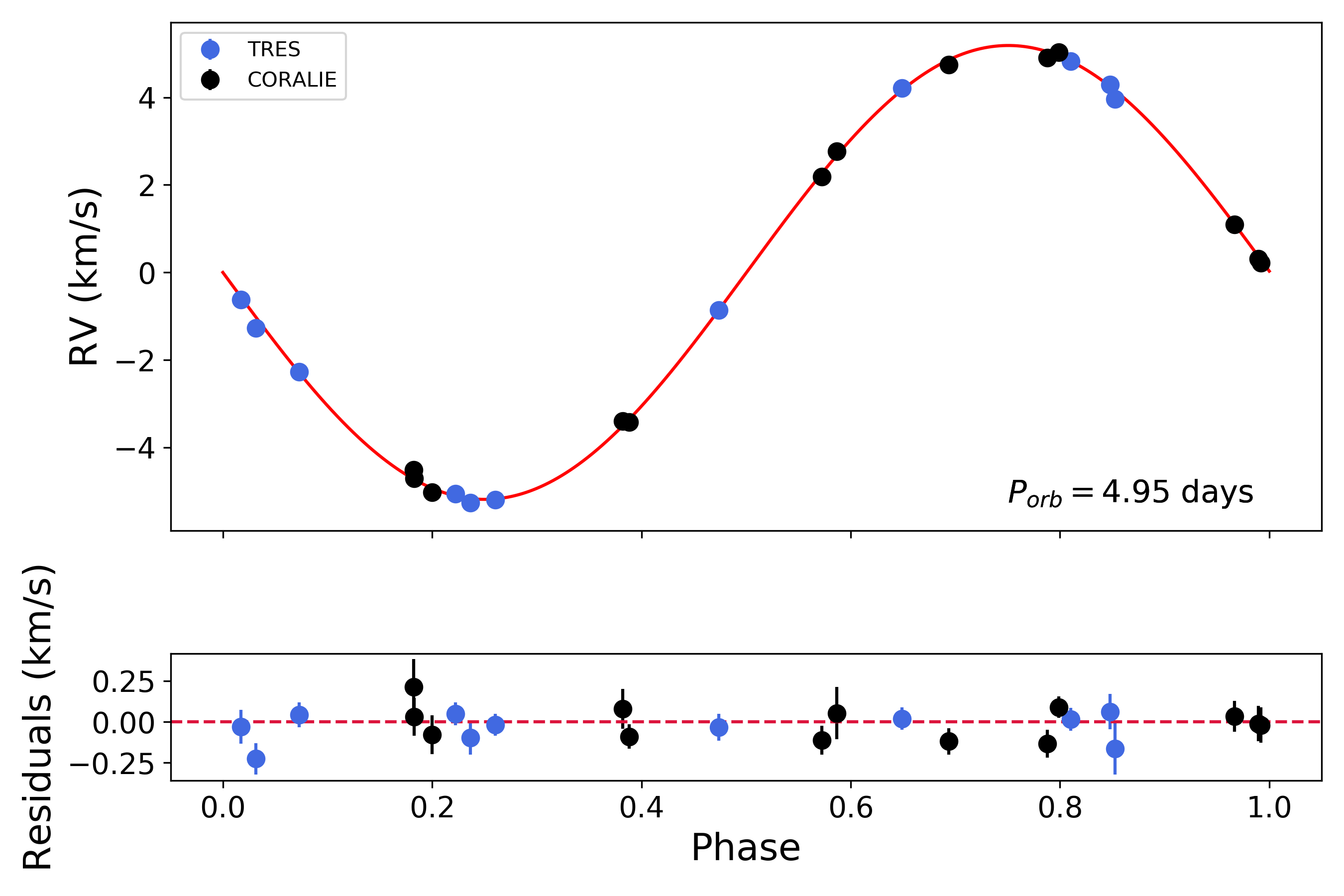}
\caption{Top: TESS and ULMT light curves of TOI-852 with global {\tt EXOFASTv2} transit model in red. The binning for the TESS data uses bins averaged over 150 minutes. Bottom: TRES and CORALIE multi-order relative radial velocities of TOI-852 with {\tt EXOFASTv2} orbital solution plotted in red.}\label{fig:toi852_obs}
\end{figure}

\section{Analysis}\label{sec:analysis}
\subsection{Modeling with {\tt EXOFASTv2}}\label{sec:exofast}

We perform a global analysis using {\tt EXOFASTv2} to derive the parameters for the primary stars and transiting BDs in the TOI-811 and TOI-852 systems. A full description of {\tt EXOFASTv2} is given in \cite{eastman2019}. {\tt EXOFASTv2} uses the Monte Carlo-Markov Chain (MCMC) method. For each MCMC fit, we use N=36 (N = 2$\times n_{\rm parameters}$) walkers, or chains, and run \dbf{until the fit passes the default convergence criteria for {\tt EXOFASTv2} (described in \cite{eastman2019}).}

Here we will describe our inputs into {\tt EXOFASTv2} and what parameters we obtain from each one. First, we input the stellar magnitudes referenced in Table \ref{tab:toi_obs} and use the spectroscopic $T_{\rm eff}$ from TRES as a prior on the SED model for each star. We also input parallax measurements from Gaia DR2, which are used with the SED model and an upper limit on V-band extinction \citep[$A_V$,][]{av_priors} to determine the stellar luminosity and radius. We use this stellar radius with the radius ratios obtained from our input ULMT, MEarth, and TESS transit photometry to constrain the radius of each transiting BD. These light curves also provide an inclination, which we combine with our input RV follow up using CHIRON, CORALIE, and TRES to constrain the mass and orbital properties of each transiting BD. For each different RV instrument, we let the RV offset $\gamma$ be a free parameter. Our input spectroscopic [Fe/H] and $T_{\rm eff}$ measurements from SPC are used as priors on the built-in MIST stellar isochrone models \citep{mist1, mist2, mist3} in {\tt EXOFASTv2}. \dbf{These models are detailed in these three studies, but we note here that the MIST models are fairly accurate in determining the ages of TOI-811 and TOI-852 given the parameter space for [Fe/H], $\log{g}$, and $T_{\rm eff}$ for each star.}

We set uniform $\mathcal{U}[a,b]$ or Gaussian $\mathcal{G}[a,b]$ priors on our input parameters. We use our spectroscopic measurements of [Fe/H] and $T_{\rm eff}$ and parallax measurements from Gaia DR2 to \dbf{define our Gaussian priors, which have width $b$ and mean $a$ of each parameter. \cite{eastman2019} gives a detailed description of how priors are implemented in {\tt EXOFASTv2}.} We compare the input $T_{\rm eff}$ and [Fe/H] from TRES to the same parameters from CORALIE and {\tt EXOFASTv2} in Table \ref{tab:spec_values} and find them to be reasonably consistent with one another. The SEDs derived by {\tt EXOFASTv2} for each star are shown in Figure \ref{fig:sed}. An abbreviated list of derived parameters is shown in Table \ref{tab:bdlist} with the full set of derived parameters and input priors for each system shown in Table \ref{tab:exofast_toi811} and Table \ref{tab:exofast_toi852}. The orbital solution and transit model for TOI-811 is shown in Figure \ref{fig:toi811_obs} and for TOI-852 in Figure \ref{fig:toi852_obs}.

\subsubsection{Blended light in the TOI-811 system}
Blended light from the nearby proper-motion companion (TIC 100757804, 2MASS J05520724-3255263) to TOI-811 (TIC 100757807) has one major effect on our analysis of the system. This is the dilution of the transit depth of the TESS light curve due to the field of view of each of the TESS pixels ($21\arcsec \times 21\arcsec$) and the MEarth light curve given our extraction aperture size of $8\farcs4$ in the RG715 band. This is particularly important in the case of TOI-811 since the visual proper-motion companion is relatively close to the target star at a distance of $4\farcs35$ and the secondary star is only 5 times fainter (roughly two $G$ magnitudes fainter, see Table \ref{tab:gaia}). The additional light from the proper-motion companion star combined with the light from the primary star in the TESS and MEarth light curves makes the transit appear shallower than it actually is, which results in a smaller BD radius. So, for TESS and MEarth light curves for TOI-811 only, we supply additional Gaussian input priors to {\tt EXOFASTv2} for the dilution factor $A_D$ of $\mathcal{G}[0.190, 0.1]$ for each instrument. \dbf{The dilution is calculated using the form $L2/(L1+L2)$ where $L2$ is the comoving secondary star and $L1$ is the primary star in the TOI-811 system.} The inclusion of these priors on the dilution in the light curves results in a BD radius that is $\bf 0.19{\rm R_J}$ larger than when ignoring any effects of dilution. \dbf{This relatively large change in the radius is reasonable given the photometric contribution of the comoving secondary star is $G<2$ magnitudes at a $4\farcs35$ separation (see Table \ref{tab:gaia}).}

\begin{figure}[!ht]
\centering
\includegraphics[width=0.47\textwidth, trim={0.5cm 0.0cm 1.25cm 0.5cm}]{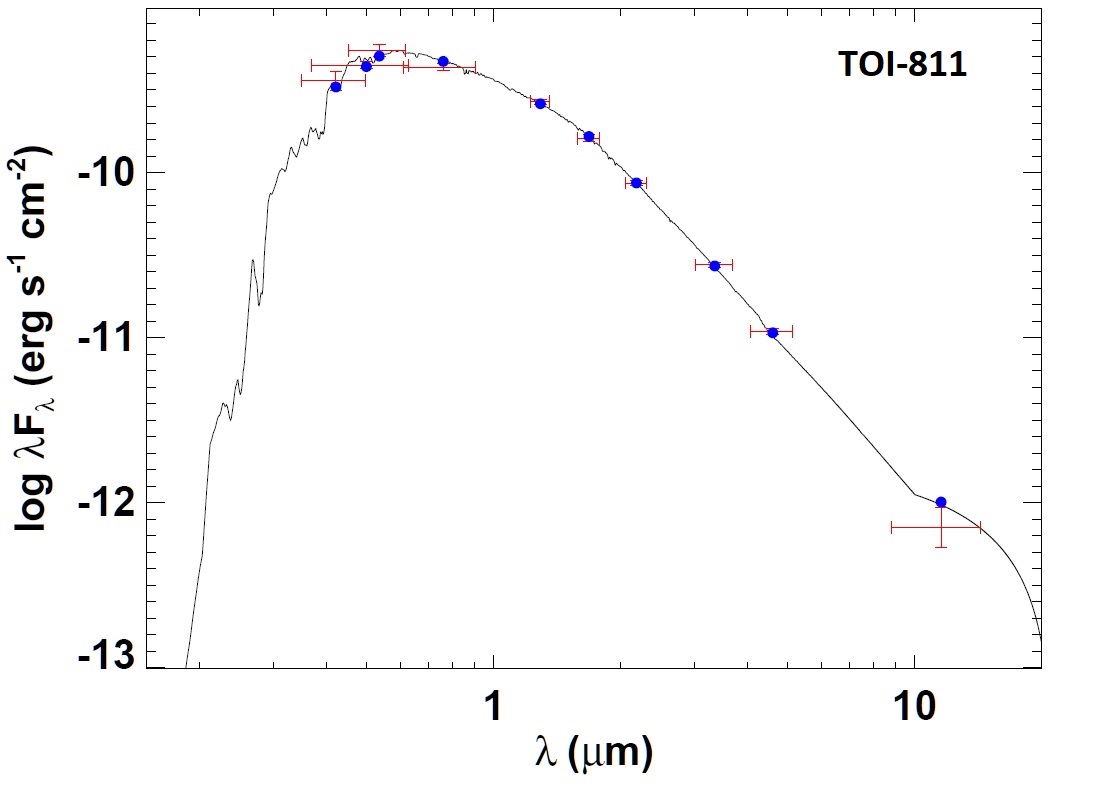}
\includegraphics[width=0.47\textwidth, trim={0.5cm 0.0cm 0.5cm 0.5cm}]{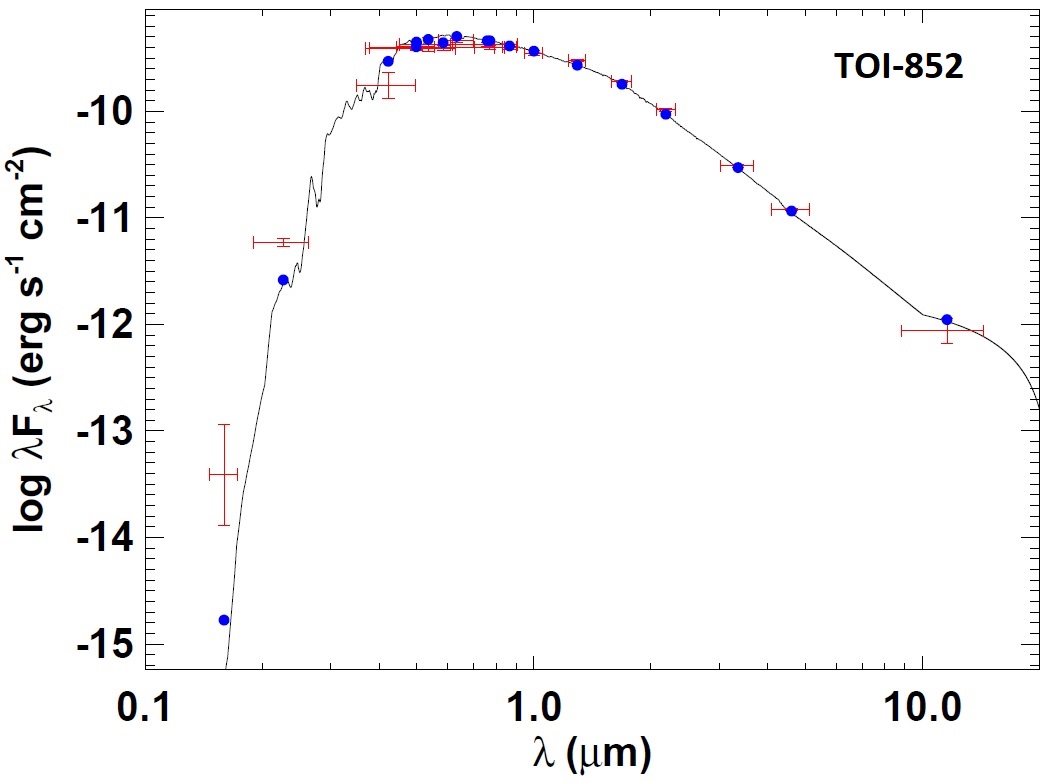}
\caption{{\tt EXOFASTv2} SEDs for TOI-811 and TOI-852. Red symbols represent the observed photometric measurements, where the horizontal bars represent the effective width of the bandpass. Blue symbols are the model fluxes from the best-fit Kurucz atmosphere model (black).}\label{fig:sed}
\end{figure}

Since we do not fit for the TESS bandpass in the SED of TOI-811, our results for the luminosity and radius of the target star are unaffected by this bandpass. The catalog photometry for the other bandpasses ($B_T$, $V_T$, $J$, $H$, $K$, WISE1, WISE2, and WISE3) resolve the target star from the neighboring star, so no deblending correction is necessary for our SED model of TOI-811 shown in Figure \ref{fig:sed}.

\subsubsection{Bimodality in the solution for TOI-852}

\begin{figure}[!ht]
\centering
\includegraphics[width=0.49\textwidth, trim={0.0cm 0.0cm 0.0cm 0.0cm}]{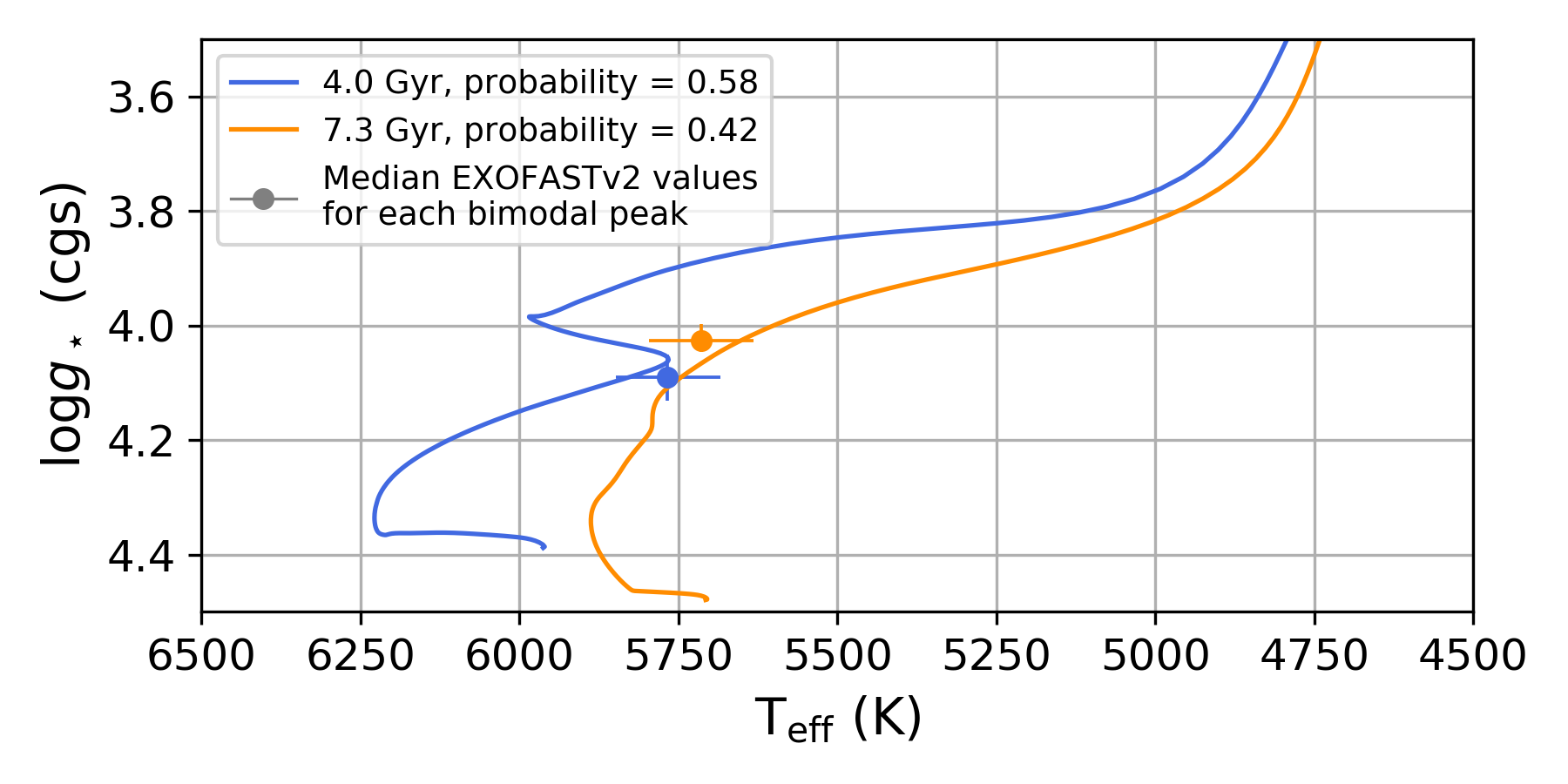}
\includegraphics[width=0.47\textwidth, trim={0.0cm 0.0cm 0.0cm 0.0cm}]{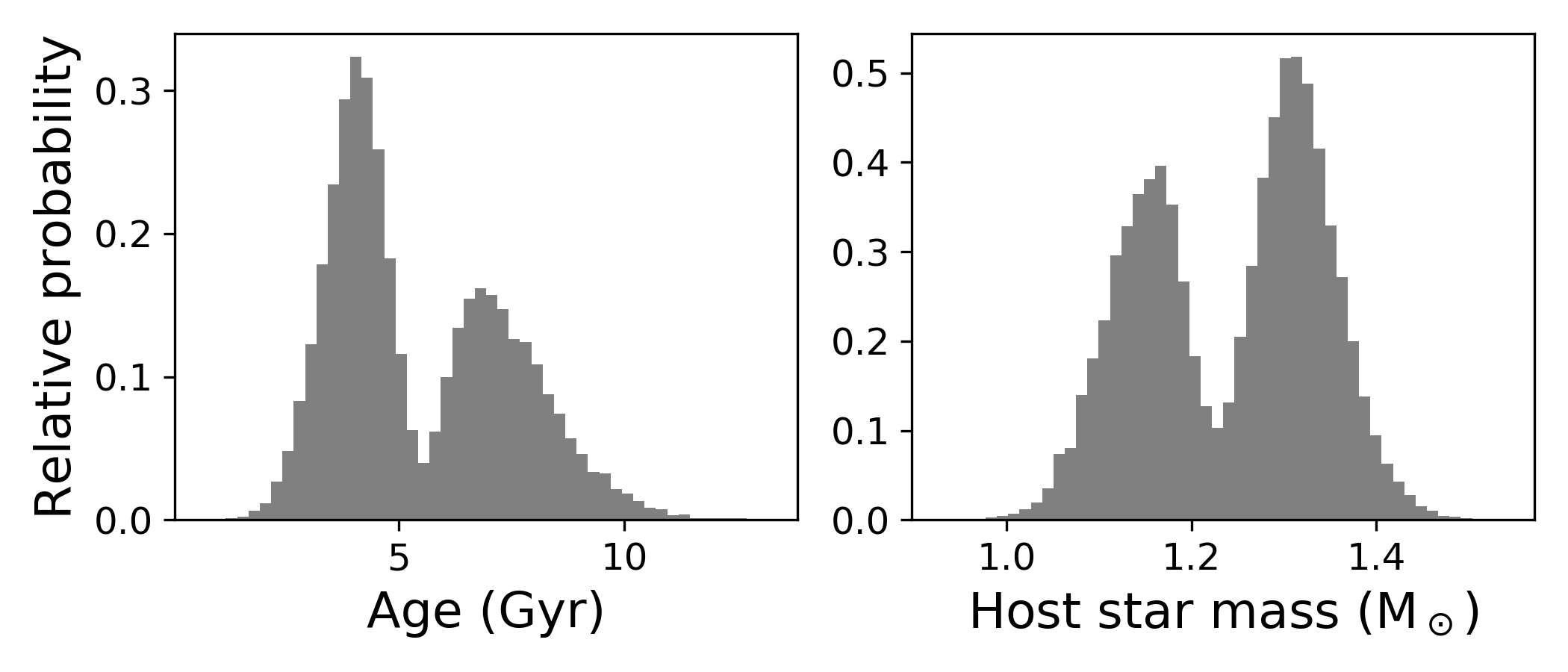}
\caption{Top: MIST isochrones from {\tt EXOFASTv2} for TOI-852. When splitting this bimodal solution, the results are shown as the blue and orange points, which show the median values and 1$\sigma$ errors for each peak. The blue color shows the higher probability solution for $\log{g}$ and $T_{\rm eff}$ and the orange shows the lower probability solution. Bottom: Age and stellar mass posterior distributions from {\tt EXOFASTv2} for TOI-852. We show these to provide a sense of the relative probabilities between the peaks of the bimodal distribution, which is in favor of a younger system ($4.0^{+0.7}_{-0.8}$ Gyr) over an older system ($7.3^{+1.3}_{-0.9}$ Gyr).}\label{fig:bimodal}
\end{figure}

We see bimodality in the posterior distribution for the age (and correlated parameters) of TOI-852, so we present the two most probable solutions resulting from the bimodal posterior distributions with the absolute most probable solution taken as the final adopted value (Table \ref{tab:exofast_toi852}). The probability is calculated based on the area of the distribution and the reported value \dbf{is the median of each probability distribution split at $M_\star=1.23{\rm M_\odot}$}. The most relevant bimodal posterior distributions are shown in Figure \ref{fig:bimodal}. The higher probability solution we report here has a probability of 0.58, with the less likely solution having a probability of 0.42. Though the differences in relative probability are marginal, we still adopt the higher of the two and simply show both solutions. This bimodality results from \dbf{the the degeneracy between the isochrones in this region of $\log{g}-T_{\rm eff}$ space that TOI-852 occupies} \dsout{the way {\tt EXOFASTv2} interpolates between the MIST isochrones}. In this case, the two curves we show in Figure \ref{fig:bimodal} are the two MIST isochrones that the MCMC analysis \dbf{found to be the best fits given this degeneracy and} the input priors on $T_{\rm eff}$ and [Fe/H] (we do not supply any priors on the surface gravity $\log{g}$). 

We see no evidence of bimodality in the posterior distributions from the MCMC analysis of TOI-811.

\subsection{Age indicators for TOI-811}
We find evidence in the Li I $\rm 6708\AA$ absorption feature and in the rotation rate and color of TOI-811 that indicate the youth of the system. We also note several other points on the activity and age of this system: 1) With the exception of the rotational spot modulation in the TESS light curves, TOI-811 is a relatively quiet young star. We find no significant Ca II H and K core emission in the TRES spectra. We also find no significant core emission in the Ca II infrared triplets. TOI-811 does not correspond with any known UV or X-ray sources, though this is expected given its distance. 2) TOI-811 does not belong in any known clusters or associations. A query with the BANYAN $\Sigma$ tool \citep{gagne2018} could not place TOI-811 into any well characterized associations. We perform a kinematics search for comoving members that may be associated with it. We queried a $25^\circ$ region about TOI-811 using the Gaia DR2 catalog for stars brighter than $R_p < 13$, and with distances within 50\% of that of the target star. This target list is cross matched with X-ray sources from ROSAT to identify any potential young stars that may be active. We then retrieve the TESS full-frame image light curves of any potential young stars that cross matches within this 3D box about TOI-811, and manually vet them for activity signatures. We find 39 stars within the region around TOI-811 that show strong rotational modulation and X-ray emission. However, they exhibit no structure in their UVW velocity distribution, and the velocities of TOI-811 did not correspond with any other stars within this subsample. In addition, this subsample does not form a distinct co-evolving population based on their colors, magnitudes, and rotation periods. As such, we conclude that we cannot identify any comoving members that are associated and coevolving with TOI-811. 

\subsubsection{Lithium abundance in TOI-811}
One indicator of age in young stars is the abundance of lithium \citep{lambert_2004}. \dbf{This is because the star reaches sufficient temperatures to destroy the lithium after it reaches a certain evolutionary state (age), so the presence of lithium implies that the star has not reached this state yet.} We use a co-added TRES spectrum of TOI-811 to measure the equivalent width (EW) of the Li I $\rm 6708\AA$ absorption line to be $\rm 0.133 \pm 0.024\AA$ following \cite{zhou2020}. We estimate the Li I absorption strength by simultaneously fitting a set of Gaussian profiles to the region around $\rm 6708 \,\AA$. Gaussian profiles with centroids at $\rm 6707.76\,\AA$ and $\rm 6707.91\,\AA$ account for the Li doublet while Gaussian profiles at $\rm 6707.43\,\AA$ account for the blended Fe I line. We assume that each line has equal width, and the two Li doublet lines have equal heights. We compare the Li EW to empirical relationships of EW and $T_{\rm eff}$ of stars in clusters and associations \citep[e.g.][]{somers2015, fang2018} in Figure \ref{fig:lithium}, and from this we find that TOI-811 has an age not inconsistent with the stars in Pleiades (125 Myr) or IC 2602 and IC 2391 (30-50 Myr). Though we do not explicitly use the Li $\rm 6708\AA$ EW measurement to constrain the age of TOI-811 in our analysis, we may qualitatively use it to confirm that TOI-811 is likely younger than 200 Myr. There is no detectable evidence of a lithium absorption feature in the TRES spectra of TOI-852.

\subsubsection{\bf Youth indicators for TIC 100757804}
\begin{figure}[!ht]
\centering
\includegraphics[width=0.48\textwidth, trim={0.0cm 0.0cm 0.0cm 0.0cm}]{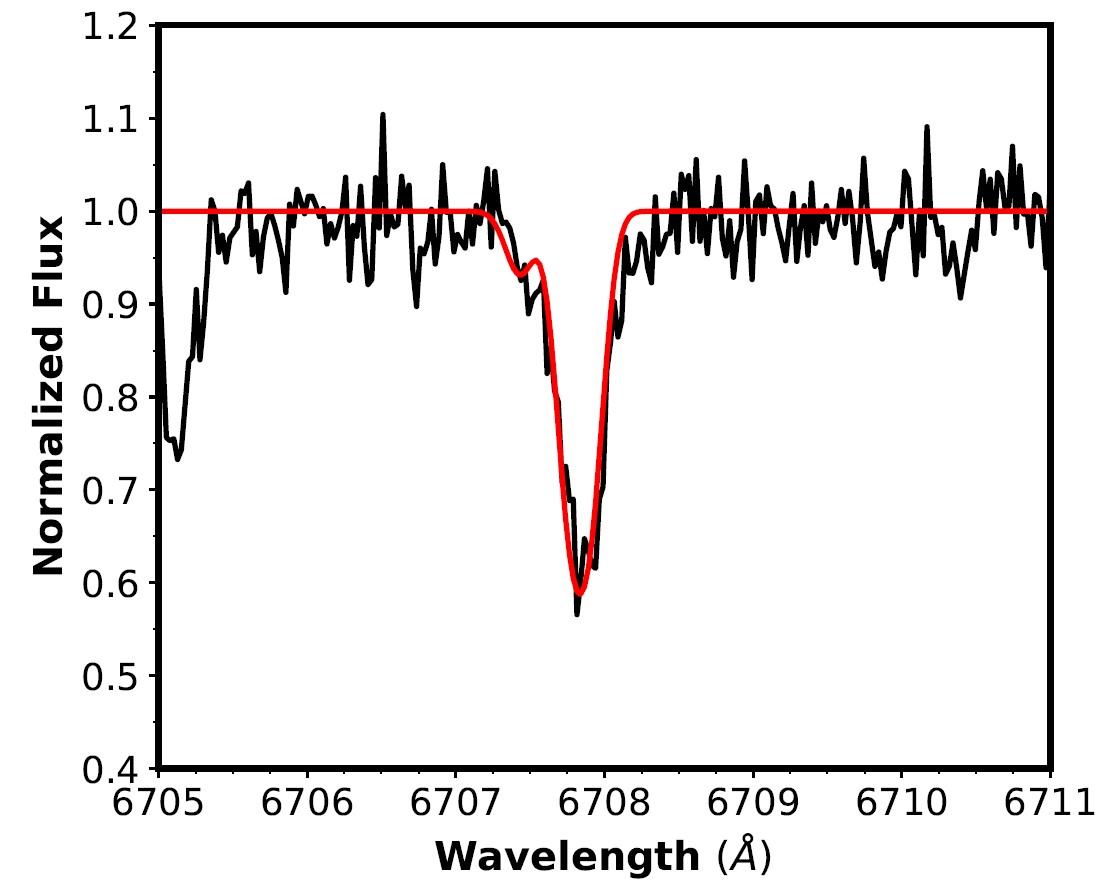}
\includegraphics[width=0.48\textwidth, trim={0.0cm 0.0cm 0.0cm 0.0cm}]{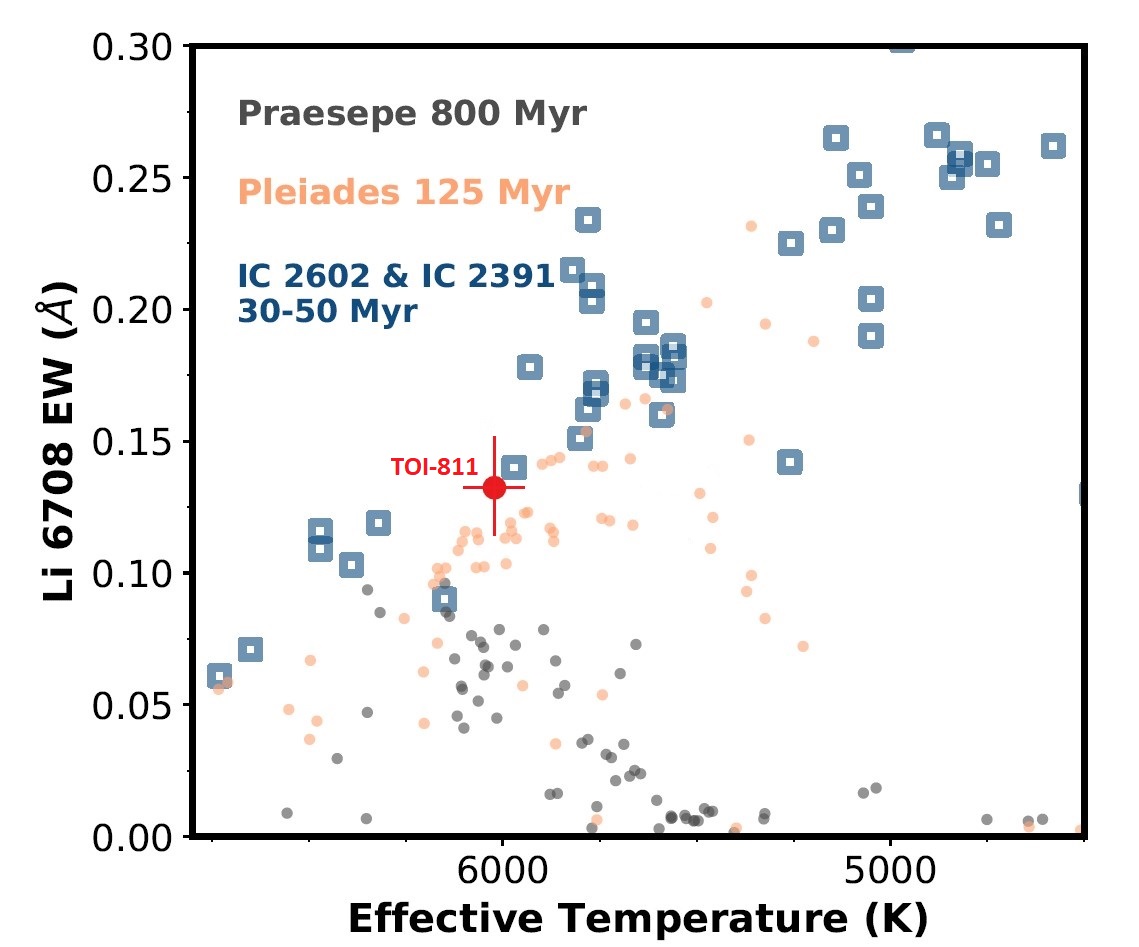}
\caption{Top panel: Normalized TRES spectrum of TOI-811 centered on the lithium absorption line. The model fit to the line and used to measure EW is shown in red. Bottom panel: Equivalent width versus effective temperature showing TOI-811 in red and several stellar clusters and associations at a range of different ages. See \citet{stauffer_1997, bouvier_2018} for lithium EW and stellar $T_{\rm eff}$ values in these cluster.}\label{fig:lithium}
\end{figure}

\dbf{The Goodman spectra described in Section \ref{sec:goodman} were made in an effort to measure additional age indicators in the spectrum of TIC 100757804. We found no significant Li $\rm 6708\AA$ (EW$\rm <0.01\AA$) feature and an H-alpha (H$\alpha$) of $\rm EW = 0.25\AA$. This lack of Li $\rm 6708\AA$ indicates that the system is likely not younger than the Pleiades (125 Myr; see \cite{somers2015} for Li $\rm 6708\AA$ abundances in Pleiades stars) but the presence of H$\alpha$ implies that the system is also not older than Praesepe (600-800 Myr). This is consistent with the youth indicators for TOI-811 that we explore in and highlight in Figure \ref{fig:lithium}.}

\subsubsection{Gyrochronology of TOI-811}\label{sec:gyro}
For the first time, we are able to use gyrochronology of a host star to constrain the age of its transiting BD. Gyrochronology takes the ideas of the Skumanich relationship and combines it with the $B-V$ magnitude (as a proxy for stellar mass) of stars to estimate the stellar age. A formulation for estimating the age via a ``gyrochrone" was developed by \cite{barnes_2007} and further refined by \cite{mamajek_hillenbrand_2008} to accommodate a wider range of stellar ages. Here we use the refined formulation by \cite{mamajek_hillenbrand_2008} in the following equation:

\begin{equation}\label{eq:gyro}
    P_{\rm rot}(B-V,t) = a\left[(B-V)-c\right]^bt^n
\end{equation}

\noindent where the stellar rotation period $P$ is a function of the $B-V$ color and age $t$ of the star, $a=0.407 \pm 0.021$ and $b=0.325 \pm 0.240$ are constants, $c=0.495 \pm 0.010$ is the ``color singularity", and $n= 0.566 \pm 0.008$ is the time-dependent power law. The rotation period $P$ is given in days and the stellar age $t$ is given in Myr. For TOI-811, we seek to determine the stellar age using the rotation period $P_{\rm rot} = 3.21 \pm 0.02$ days and the $B-V=0.707\pm 0.203$ color (from Table \ref{tab:toi_obs}), so we solve for $t$ in Equation \ref{eq:gyro}. This yields an estimate for the age of TOI-811 to be $93^{+61}_{-29}$ Myr, which is consistent with the upper limit to the age of 200 Myr indicated by the Li $\rm 6708\AA$ EW measurement.

\dbf{We note here that even though the \cite{mamajek_hillenbrand_2008} model for gyrochronology yields a relatively well-determined estimate for the age of TOI-811, in general, there is a larger uncertainty in this age determination depending on the choice of coefficients in Equation \ref{eq:gyro}. This idea can be demonstrated when choosing different coefficients (from \cite{angus2019}), where the age of TOI-811 would be $230\pm 30$ Myr instead of $93^{+61}_{-29}$ Myr. That said, our confirmation of the youth of the TOI-811 system remains unchanged.}

Although we could apply the same analysis to TOI-852, gyrochronology is best applied to young solar analogs and especially those without close-in massive companions, which may raise tides in the surface of the star and affect the stellar rotation period. Young solar analogs have more surface activity in the form of star spots, for example, and so are easier to extract a rotation period from with a periodogram analysis of their light curves. Though we are able to relatively easily obtain a rotation period from the modulation in the light curve of TOI-852, we find two disqualifying features about the system that discourage a pursuit of a gyrochronologic measurement. 

The first is that we find a bimodal age distribution that indicates that the host star is \dsout{old} \dbf{evolved} (4 Gyr or 7 Gyr) based on the MIST models. Though the bimodal nature of the distribution in age makes a precise age determination more difficult, we can confidently conclude that TOI-852 is an \dsout{old} \dbf{evolved} system. \dbf{Evolved solar analog stars are especially poor candidates for gyrochronology given the large spread in their rotation periods \citep{vanSaders2019}. In other words, though it may be the case that we can constrain the age of an evolved or subgiant star relatively well, that will not be useful in a gyrochronologic analysis given subgiants have rotation periods ranging from 1 to 50 days as \cite{vanSaders2019} show.}

The second disqualifying feature is that the BD companion has likely had non-negligible tidal influence on the rotation period of the star given the system's mass ratio and short orbital period.

\subsection{Rotational inclination angles of TOI-811 and TOI-852}\label{sec:incl}

\begin{figure}[!ht]
\centering
\includegraphics[width=0.49\textwidth, trim={0.0cm 0.0cm 0.0cm 0.0cm}]{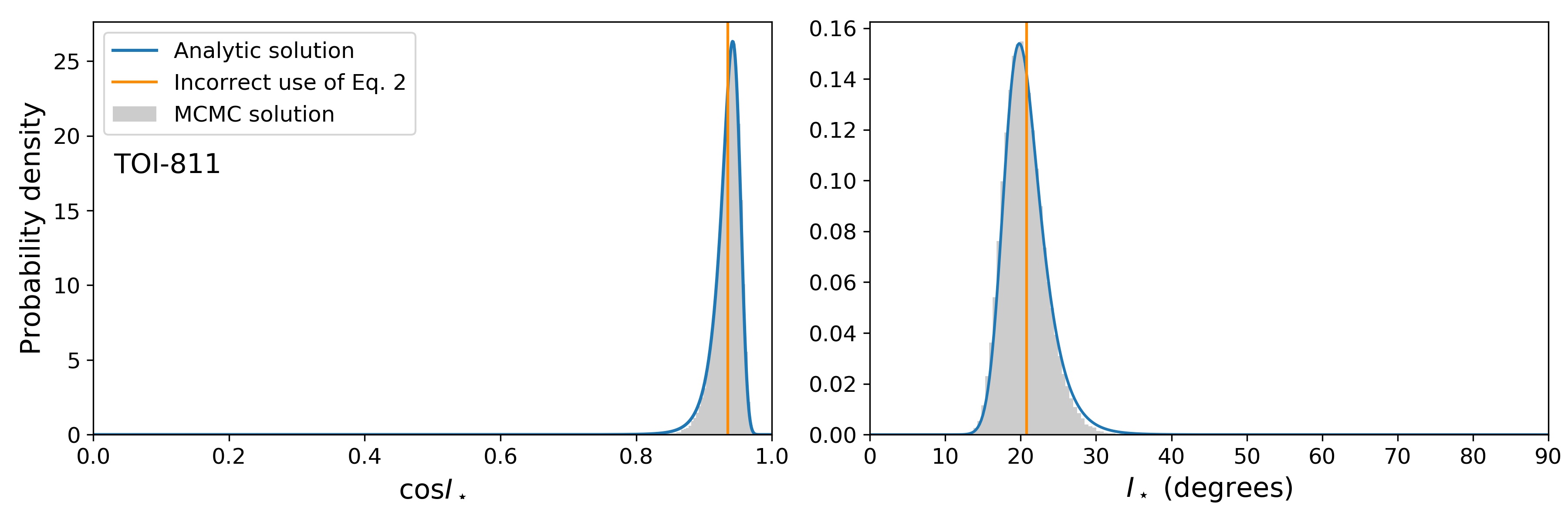}
\includegraphics[width=0.49\textwidth, trim={0.0cm 0.0cm 0.0cm 0.0cm}]{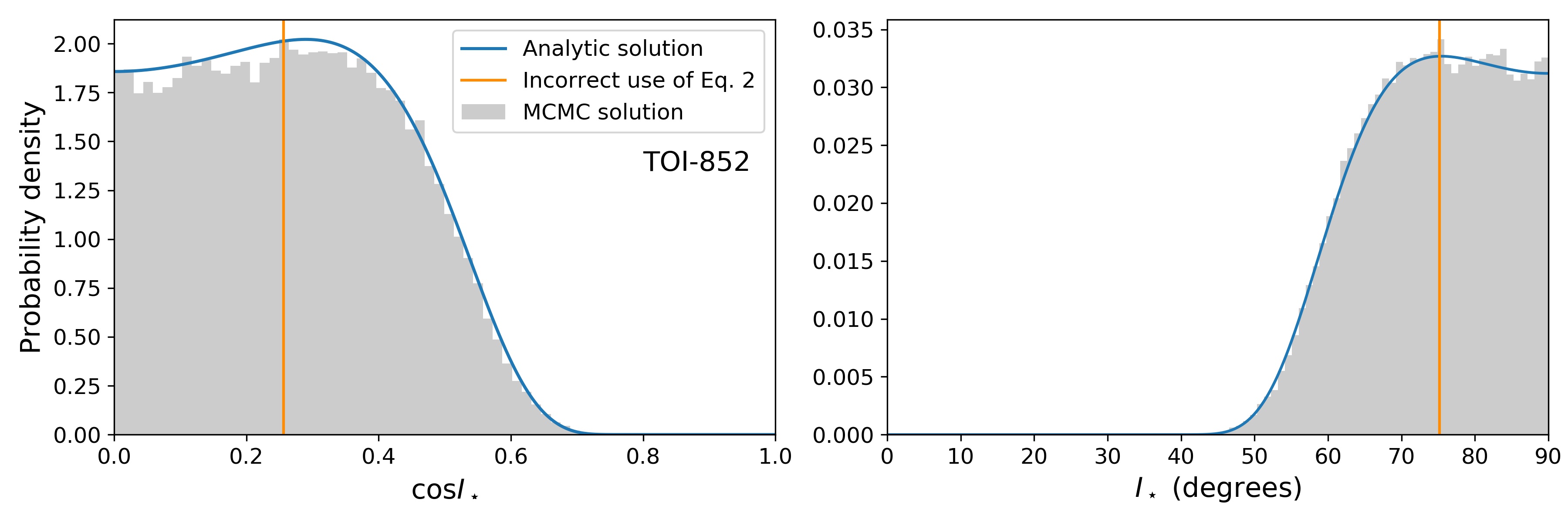}
\caption{Probability distributions of $\cos{I_\star}$ and $I_\star$ for TOI-811 (top) and TOI-852 (bottom). The analytic and MCMC solutions follow the procedure outlined by \cite{masuda2020}. TOI-811 has an inclination angle $I_\star = 19.62^{+3.62}_{-1.81}\,^\circ$, which is misaligned and nearly polar with the orbit of the BD and to the equator of the stellar spin. In contrast, TOI-852 has an inclination angle $I_\star = 73.12^{+11.88}_{-9.86}\,^\circ$, which is only marginally misaligned (near the equator of the stellar spin) with the orbit of TOI-852b. For completeness, we show that the incorrect use of Equation \ref{eq:incl} results in no information about the asymmetry of the uncertainties and that it differs slightly from the peak of the analytic solution. \label{fig:incl}}
\end{figure}

Whenever we detect photometric modulation in the light curve of a star, we have the opportunity to combine the information about the stellar rotation period within that modulation with the projected equatorial velocity of the star to determine the inclination angle $I_\star$. When compared to the inclination angle of a transiting object, in our case transiting BDs, then we may learn of the host star obliquity. The traditional way of calculating $I_\star$ is:

\begin{equation}\label{eq:incl}
    I_\star = \sin^{-1}\left(\frac{v\sin{I_\star}}{V_{\rm rot}}\right)   
\end{equation}

\noindent where our observed $v\sin{I_\star}$ values come from the TRES spectra of each star and $V_{\rm rot} = 2\pi R_\star/P_{\rm rot}$. We obtain $R_\star$ from our {\tt EXOFASTv2} results and $P_{\rm rot}$ from our Lomb-Scargle periodogram analysis for each star. Although Equation \ref{eq:incl} appears quite simple, in practice we must be cautious in our use of it. Because the priors on $v\sin{I_\star}$ and $V_{\rm rot}$ are dependent on each other, we must use statistical inference to calculate $I_\star$, otherwise we will bias the resulting $I_\star$ and lack information on the uncertainties on that measurement. By leveraging the probability distributions of $R_\star$ and $v\sin{I_\star}$ properly, we produce a distribution of $\cos{I_\star}$ and $I_\star$ values from which we derive uncertainties for the stellar inclination. The details of this treatment are outlined in \cite{masuda2020} and our results using their procedure are shown in Figure \ref{fig:incl}. For TOI-811, $v\sin{I_\star}= 7.11\pm 0.50\, {\rm km\,s^{-1}}$ and $V_{\rm rot} = 19.44^{+1.35}_{-2.44}\, {\rm km\,s^{-1}}$. For TOI-852, $v\sin{I_\star}= 14.50\pm 0.50 \,{\rm km\,s^{-1}}$ and $V_{\rm rot} = 14.97\pm 0.34\, {\rm km\,s^{-1}}$.

TOI-811 has an inclination angle $I_\star = 19.62^{+3.62}_{-1.81}\,^\circ$ and when compared to the orbital inclination angle $i=89.56^{+0.28}_{-0.24}\,^\circ$ of the BD, we clearly see that the TOI-811 system is misaligned. TOI-852 has an inclination angle $I_\star = 73.12^{+11.88}_{-9.86}\,^\circ$ and this is comparable to the orbital inclination angle $i=84.74^{+0.28}_{-0.27}\,^\circ$ of the BD in this system. This means that TOI-852 is marginally misaligned. Additionally, without a measurement of the obliquity of the orbital plane of the transiting BDs (for example, from in-transit Doppler tomography measurements), we only have the line-of-sight-projected orbital obliquity. It is important to note that these are projected alignments and that the stellar inclination angle may be offset by $180^\circ$ and still show the same relative alignment to the orbit of the BD. 

\begin{figure}[!ht]
\centering
\includegraphics[width=0.49\textwidth, trim={1.0cm 1.0cm 0.0cm 1.0cm}]{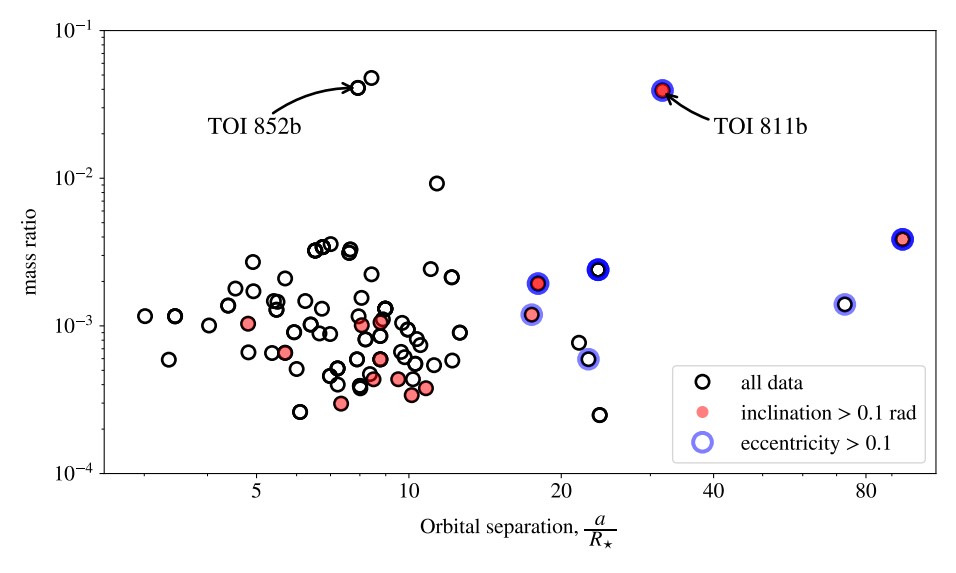}
\caption{Mass ratio versus orbital separation ($a/R_\star$) for a representative sample of giant planets orbiting stars of $T_{\rm eff}<6250$K \citep{tepcat} with TOI-852b and TOI-811b highlighted. Though both BDs are outliers in mass, they follow the pattern seen in their giant planet counterparts: circular, aligned orbits for $a/R_\star<8$ or eccentric, misaligned orbits for $a/R_\star>8$. \label{fig:obliq}}
\end{figure}

\dbf{We show in Figure \ref{fig:obliq} where TOI-811b and TOI-852b appear in the context of giant planets for which we have information on the obliquity and eccentricity, and for which the host star has $T_{\rm eff}< 6250$K. These BDs are consistent with the pattern seen in the giant planet population where objects with $a/R_\star < 8$ are in well-aligned, circular orbits and those $a/R_\star > 8$ are eccentric and sometimes misaligned \citep{albrecht2012}.}

\section{Discussion}\label{sec:conclusion}
With the discoveries of TOI-811b and TOI-852b, we add two more transiting BDs with well-determined masses and radii (i.e. with uncertainties around the 5-10\% level) to the transiting BD population. Two years ago, we knew of only 17 transiting BDs (and 2 BDs in an eclipsing binary system discovered by \cite{2M0535}) and only 14 of those with uncertainties on their radius at less than 10\%. Now, we know of 25 BDs that transit a star \dbf{(and 3 additional BDs in a triple system discovered by \cite{bd_triple})} and as we show in Table \ref{tab:bdlist}, the TESS mission has made a steady contribution to the number of known transiting BDs over the span of its primary mission. Five of six of these TESS BDs also have improved precision on the measurements of their radii thanks to improved parallaxes from Gaia DR2.

TESS has covered most of the sky with at least roughly 28 days of continuous observation, which has led to discoveries of transiting BDs in orbital periods on the order of 10 days or less (to enable the detection of multiple transits). As TESS lengthens its effective coverage beyond a minimum of 28 days during its first extended mission, we will discover more transiting BDs in longer periods that are potentially similar to the likes of TOI-811b. Young transiting BDs like TOI-811b are particularly valuable since these occupy a place in the mass-radius diagram where radius changes quickly with age.

\begin{figure}[!ht]
\centering
\includegraphics[width=0.48\textwidth, trim={1.0cm 0.0cm 0.0cm 0.0cm}]{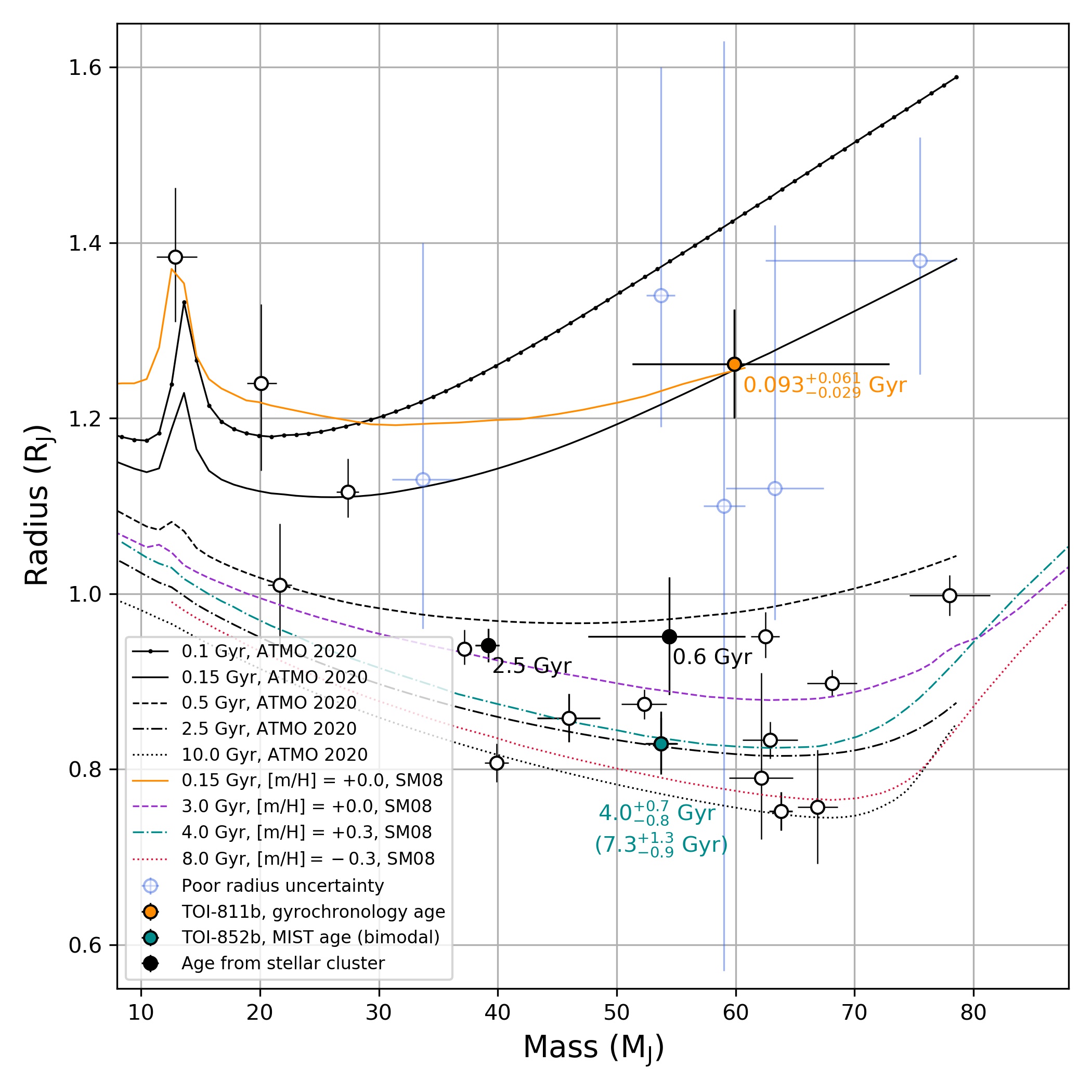}
\caption{Mass-radius diagram of transiting BDs featuring the SM08 and ATMO 2020 models. TOI-811b ($0.093^{+0.061}_{-0.029}$ Gyr) and TOI-852b ($4.0^{+0.7}_{-0.8}$ Gyr) are shown as an orange point and a cyan point, respectively. Only 3 BDs that transit main sequence stars have ages constrained by stellar clusters or associations (AD 3116b at 600 Myr in Praesepe, CWW 89Ab at 2.5 Gyr in Ruprecht 147, and RIK 72b at 10 Myr in Upper Scorpius). \dbf{We give some attention to CWW 89Ab (labelled by ``2.5 Gyr") to show that the SM08 models can demonstrate consistency with it, but refer to \cite{cww89a} for a detailed review of this system.} Note RIK 72b is not shown because its radius is $3.1\rj$. Also not shown are the BD binary systems, 2MASS J05352184–0546085 (located in the Orion Nebula Cluster with an age of 1-2Myr) \dbf{and 2MASS J1510478-281817 (in the Argus moving group, 45 Myr) given that these are not transiting stars}. We also update the radius of AD 3116b using the Gaia DR2 parallax of its host star for this figure.} \label{fig:massrad}
\end{figure}

As we discover more transiting BDs, we must generally give more statistical weight to those with well-determined masses, radii, and ages. Transiting BDs with radius uncertainties above the 10\% level (such as CoRoT-33b and TOI-503b) though valuable for the census of the transiting BD population, offer little in the way of providing a firm test point for substellar isochrones \cite[e.g.][]{chabrier2000, baraffe03, saumon08, ATMO2020}. This is because transiting BDs change rapidly in radius at young ages up to 1 Gyr and they appear to asymptotically approach a minimum radius at older ages. So, a poor constraint on the radius provides little if any constraint on which substellar isochrone at a given age best matches. CoRoT-33b is a particularly bad offender in this regard given its radius uncertainties span from $\rm 0.57R_J$ to $\rm 1.63R_J$ (corresponding to the 100 Myr to beyond the 10 Gyr isochrone tracks) in the mass-radius diagram (Figure \ref{fig:massrad}).

If instead we are fortunate enough to discover a transiting BD like TOI-811b, where we have precise measurements of the BD's radius and a decent constraint on the host star's youth, we must take advantage of and give more weight to such discoveries. We should similarly treat objects like TOI-852b, where we have a precise measurement on the radius and a fair constraint on the old age (from stellar isochrone models), which reveal the BD to trace out the oldest substellar isochrones along with several other old transiting BDs.

\subsection{Testing young and old substellar isochrones}
With the new transiting BD discoveries in this study, we have the opportunity to test some of the youngest and the oldest substellar isochrones in the mass-radius diagram. We show both the ATMO 2020 and the \dbf{\cite{saumon08} (SM08)} models in Figure \ref{fig:massrad} to examine how well each describes our young and old transiting BDs in this study. \dbf{We choose to include the SM08 models to show an independent set of evolutionary tracks from the ATMO 2020 models and to highlight how metallicity affects BD evolution. However, it is important to note that we do not have any measure of the atmospheric composition of either TOI-811b or TOI-852b, so the comparison to the SM08 models assumes that the metallicity matches that of the host star. It has been theorized by works like \cite{burrows2011} that a higher metallicity causes a higher BD atmospheric opacity and that this results in the BD maintaining more internally generated heat and thus a larger radius at a given age. We do not fully explore this idea in this work, but we highlight the effects of metallicity by showing a sample of the SM08 models that we find to most closely fit TOI-811b and TOI-852b in Figure \ref{fig:massrad}.} \dsout{and how they compare to each other as only the COND03 models consider irradiation from the host star for orbital periods between 3-5 days. For ages younger than 1 Gyr, the COND03 models predict slightly larger radii for a given mass and age compared to the ATMO 2020 models. This is due in part to the fact that irradiation from the host star slows the contraction rate of the BD, but since the ATMO 2020 models use different cooling equations for the BDs, the irradiation is not the only reason for the differences seen here. Although other factors, like atmospheric metallicity in the BD, also affect the contraction rate, for the scope of this study we focus more broadly on how consistent the ATMO 2020 and SM08 models are with TOI-811b and TOI-852b.}

We find that TOI-811b occupies a region of the substellar mass-radius space with no other objects that have well-determined radii and ages to serve as tests of substellar isochrones. We have estimated the age of TOI-811b using three methods: 1) using an EW measurement of the Li $\rm 6708\AA$ absorption feature in the host star's spectrum for an age not inconsistent with $50-125$ Myr, 2) using gyrochronology and Equation \ref{eq:gyro} to estimate an age of $93^{+61}_{-29}$ Myr for the host star \dbf{(though this can have a large uncertainty as discussed in Section \ref{sec:gyro})}, 3) using the MIST isochrones (when using a prior on the age from Equation \ref{eq:gyro}) for an age estimate of $117^{+43}_{-37}$ Myr of the host star. This is all under the assumption that the host star and transiting BD formed at the same time and so are the same age. From this analysis, we are confident that the age of TOI-811b is \dbf{roughly 100-200 Myr}, thus making it a meaningful test point for young substellar isochrones and showing that the age is consistent with the \dbf{SM08 and ATMO 2020} evolutionary tracks for ages between roughly 100 and 200 Myr.

\dsout{In addition to TOI-811b, the other young transiting BDs with well-determined ages include AD 3116b (in Praesepe, 600 Myr), RIK 72b (in Upper Scorpius, 5-10 Myr), and 2MASS J05352184–0546085 (in the Orion Nebula, 1-2 Myr).}

TOI-852b is a significantly older transiting BD. Our analysis yields a bimodal result for the age of TOI-852 to be either 4.0 Gyr or 7.3 Gyr based on the MIST isochrones, which places its companion BD among the oldest known transiting BDs. The 4.0 Gyr solution is favored by a probability of 0.58 to the probability of 0.42 of the 7.3 Gyr solution. TOI-852b's age is \dbf{consistent with the 4-8 Gyr SM08 models. This may be a result of the relatively high metallicity ($\rm [Fe/H]=+0.33$) increasing the atmospheric opacity and limiting the rate of heat loss and contraction of the BD, which is a phenomenon detailed in works like \cite{saumon08} and \cite{burrows2011}. In other words, a lower metallicity yields a smaller BD at a given age and a higher metallicity yields a larger BD at that same given age. However, pay careful attention to the key in Figure \ref{fig:massrad} and note that the 4-8 Gyr SM08 models also span the metallicity range of $-0.3$ to $+0.3$ dex, which means that TOI-852b is consistent with the 8 Gyr, $-0.3$ dex SM08, if only marginally. So, the MIST models for the age of the host star in the TOI-852 system slightly favor a 4 Gyr solution and the SM08 models show a consistency with an age of 4-8 Gyr, but over a large range of metallicities.} \dsout{only marginally consistent with the 5-10 Gyr ATMO 2020 models and interestingly, TOI-852b joins the likes of KOI-205b, EPIC 2017002477b, and TOI-569b as the transiting BDs that trace out the oldest 10 Gyr mass-radius isochrones shown in Figure \ref{fig:massrad}. This may suggest that transiting BDs do not contract any smaller than roughly $R_b = 0.75 {\rm R_J}$ and that the contraction approaches this asymptotic limit as early as 4 or 5 Gyr in the lifetime of the transiting BD.}

\subsection{Orbital properties of the TOI-811 and TOI-852 systems}
We see that both TOI-811b and TOI-852b add new test points for young and old substellar isochrones, respectively. We now examine each transiting BD's orbital properties and compare them to each other. TOI-811b is a young transiting BD in an eccentric $e=0.41$ orbit in a period of $P_{\rm orb} = 25.2$ days. Based on the projected alignment of the stellar inclination angle ($I_\star = 19.62^{+3.62}_{-1.81}\,^\circ$) of TOI-811 with the orbital inclination angle ($i=89.56^{+0.28}_{-0.24}\,^\circ$) of TOI-811b, we show that the system is highly spin-orbit misaligned, with one pole of the star nearly pointing along the line of sight of the observer.

In contrast, TOI-852b orbits its host star in $P_{\rm orb}=5.0$ days and has an eccentricity consistent with zero (a circular orbit). The orbital period is nearly synchronized with the rotational period of the host star ($P_{\rm rot}=5.8$ days) and from this $P_{\rm rot}$, we are able to determine that the projected stellar inclination angle $I_\star = 73.12^{+11.88}_{-9.86}\,^\circ$ and the orbital inclination angle $i = 84.74^{+0.28}_{-0.27}\,^\circ$ of the BD are consistent with alignment. Altogether, this means that we cannot rule out that this system is old enough to have had the BD circularize in its orbit, nearly synchronize with the rotation rate of the star, and nearly align with the spin axis of the star via tidal interactions. However, it is also possible that the BD instead formed at or near its current orbital distance from the star and slowly migrated inward. This contrasts the TOI-811 system in which we have a much younger BD in a wider, eccentric orbit that is clearly misaligned with the projected spin axis of the star and not synchronized \dsout{or pseudo-synchronized} with the stellar rotation rate of $P_{\rm rot}=3.2$ days. \dbf{Though we do not explore these tidal interactions in much more detail in this study, we still note their potential significance in these systems and how they compare to others (Figure \ref{fig:obliq}).}

\subsection{Summary}
We use data from the TESS mission, ground based RV follow up, and parallax measurements from Gaia DR2 to characterize two newly discovered transiting BDs, TOI-811b and TOI-852b. We find evidence in the Li $\rm 6708\AA$ equivalent width in the spectrum of TOI-811 that the star is roughly between the ages of 50-125 Myr and that this is not inconsistent with the age of $93^{+61}_{-29}$ Myr that we calculate from gyrochronology relations \cite{mamajek_hillenbrand_2008} and with the age of $117^{+43}_{-37}$ Myr that we estimate using the MIST models. Assuming the age of the transiting BD, TOI-811b, is the same as its host star, we find that the \dbf{SM08 \citep{saumon08}} and ATMO 2020 \citep{ATMO2020} substellar isochrones are consistent with the mass, radius, and age of this transiting BD.

Our analysis of the TOI-852 system shows a bimodal age distribution that is marginally in favor of a 4.0 Gyr old system against a 7.3 Gyr old system. This system contrasts the TOI-811 system not only in age and radius difference but also in the orbital properties of the transiting BD. TOI-852b has a circular, 5-day orbit (compared to the eccentric, 25-day orbital period of TOI-811b) that is nearly synchronized with the rotation period of the host star and marginally misaligned with the spin axis of the star (contrasting the misaligned orbit of the TOI-811 system). The ATMO 2020 substellar isochrones slightly \dbf{underestimate} the age of TOI-852b, \dbf{but the SM08 models are consistent within $1\sigma$ of the 4.0 Gyr age of assumed from the MIST models of the host star.} \dsout{but this is in a region of the substellar mass-radius diagram where the isochrones between 3 and 10 Gyr begin to crowd together as the contraction of transiting BDs decelerates.}

These two new transiting BDs give us a look at young and old substellar objects and, in the case of TOI-811b, allow us to test substellar evolutionary models with a young transiting BD. For the first time, we use gyrochronology of a young star that hosts a transiting BD to benchmark substellar isochrones and we see that TOI-811b is consistent with the isochrones for transiting BDs between roughly 100 and 200 Myr in age. Though we have gradually increased the population size of transiting BDs with TESS over the past two years, we must especially pursue young transiting BDs like TOI-811b. By doing so, we create more opportunities to apply established stellar age dating methods like gyrochronology to transiting substellar objects.

\section{Acknowledgements}
Funding for the TESS mission is provided by NASA's Science Mission directorate. This paper includes data collected by the TESS mission, which are publicly available from the Mikulski Archive for Space Telescopes (MAST). Resources supporting this work were provided by the NASA High-End Computing (HEC) Program through the NASA Advanced Supercomputing (NAS) Division at Ames Research Center for the production of the SPOC data products.

This work has made use of data from the European Space Agency (ESA) mission Gaia (\url{https://www.cosmos.esa.int/gaia}), processed by the Gaia Data Processing and Analysis Consortium (DPAC, \url{https://www.cosmos.esa.int/web/gaia/dpac/consortium}). Funding for the DPAC has been provided by national institutions, in particular the institutions participating in the Gaia Multilateral Agreement.

Funding for this work is provided by the National Science Foundation Graduate Research Fellowship Program Fellowship (GRFP).

TWC acknowledges the efforts of the members of the TESS Follow up Program and the Science Processing Operations Center in making the TESS data readily accessible for the analysis in this work. TWC also thanks Adam Kraus and Elisabeth Newton for contributing discussions.

The MEarth Team gratefully acknowledges funding from the David and Lucile Packard Fellowship for Science and Engineering (awarded to D.C.). This material is based upon work supported by the National Science Foundation under grants AST-0807690, AST-1109468, AST-1004488 (Alan T. Waterman Award), and AST-1616624, and upon work supported by the National Aeronautics and Space Administration under Grant No. 80NSSC18K0476 issued through the XRP Program. This work is made possible by a grant from the John Templeton Foundation. The opinions expressed in this publication are those of the authors and do not necessarily reflect the views of the John Templeton Foundation.

This work makes use of observations from the LCOGT network.

\dbf{The authors thank the reviewer for the timely feedback and thoughtful comments.}

\facilities{TESS, Las Cumbres Observatory Global Telescope (LCOGT), SuperWASP, SOAR (HRCam), Gaia, FLWO:1.5m (TRES), CTIO:1.5m (CHIRON), Euler:1.2m (CORALIE), WISE (infrared), CTIO:2MASS (optical, infrared), MEarth (Optical)} 

{\dbf \software{{\tt EXOFASTv2} \citep{eastman2019}, {\tt SPC} \citep{spc}, LCO {\tt BANZAI} \citep{Collins:2017}, {\tt AstroImageJ} \citep{Collins:2017}} }

\bibliographystyle{aasjournal}
\bibliography{citations}

\begin{deluxetable}{cccccccccc}

\tabletypesize{\footnotesize}
\tablewidth{0pt}

 \tablecaption{List of published transiting \dbf{and eclipsing brown dwarfs} as of June 2020. \label{tab:bdlist}}

 \tablehead{
 \colhead{Name} & \colhead{$P$ (days)} & \colhead{$\rm M_{BD}/M_J$} & \colhead{$\rm R_{BD}/R_J$}& \colhead{e} & \colhead{$\rm M_\star/\mst$} &\colhead{$\rm R_\star/\rst$}& \colhead{$\rm T_{eff} (K)$}&\colhead{[Fe/H]} &\colhead{Reference}}
 \startdata 
TOI-811b & 25.166 & $59.9^{+13}_{-8.6}$ & $  1.26\pm 0.06$ & $  0.509\pm 0.075$ & $1.32 \pm 0.07$ & $1.27\pm 0.09$ & $6107\pm 77$ & $+0.40 \pm 0.09$ & this work\\
 TOI-852b & 4.946 & $  53.7\pm 1.4$ & $  0.83 \pm 0.04$ & $0.004\pm 0.004$ & $1.32 \pm 0.05$ & $1.72 \pm 0.04$ & $5768\pm 84$ & $+0.33 \pm 0.09$ & this work\\
 HATS-70b & 1.888 & $12.9\pm 1.8$ & $1.38\pm 0.08$ & $<0.18$ & $1.78 \pm 0.12$ & $1.88\pm 0.07$ & $7930\pm 820$ & $+0.04 \pm 0.11$ & 1\\
 KELT-1b & 1.218 & $27.4 \pm 0.9$ & $1.12 \pm 0.04$ & $0.01 \pm 0.01$ & $1.34 \pm 0.06$ & $1.47 \pm 0.05$ & $6516 \pm 49$ & $+0.05 \pm 0.08$ & 2\\
 NLTT 41135b & 2.889 & $33.7 \pm 2.8$ &  $1.13 \pm 0.27$ & $<0.02$ & $0.19 \pm 0.03$ & $0.21 \pm 0.02$ & $3230 \pm 130$ & $-0.25 \pm 0.25$ & 3\\
 LHS 6343c & 12.713 & $62.9 \pm 2.3$ & $0.83 \pm 0.02$ & $0.056 \pm 0.032$ & $0.37\pm 0.01$ & $0.38\pm 0.01$ & - & $+0.02 \pm 0.19$ & 4\\
 LP 261-75b & 1.882 & $68.1 \pm 2.1$ & $0.90 \pm 0.02$ & $<0.007$ & $0.30 \pm 0.02$ & $0.31 \pm 0.01$ & $3100 \pm 50$ & - & 5\\
 WASP-30b & 4.157 & $62.5 \pm 1.2$ & $0.95 \pm 0.03$ & 0 (adopted) & $1.25 \pm 0.04$ & $1.40 \pm 0.03$ & $6202 \pm 51$ & $+0.08 \pm 0.10$ & 6\\
 WASP-128b & 2.209 & $37.2 \pm 0.9$ & $0.94 \pm 0.02$ & $<0.007$ & $1.16 \pm 0.04$ & $1.15 \pm 0.02$ & $5950 \pm 50$ & $+0.01 \pm 0.12$ & 7\\
 CoRoT-3b & 4.257 & $21.7 \pm 1.0$ & $1.01 \pm 0.07$ & 0 (adopted) & $1.37 \pm 0.09$ & $1.56 \pm 0.09$ & $6740 \pm 140$ & $-0.02 \pm 0.06$ & 8\\
 CoRoT-15b & 3.060 & $63.3 \pm 4.1$ & $1.12 \pm 0.30$ & 0 (adopted) & $1.32 \pm 0.12$ & $1.46 \pm 0.31$ & $6350 \pm 200$ & $+0.10 \pm 0.20$ & 9\\
 CoRoT-33b & 5.819 & $59.0 \pm 1.8$ & $1.10 \pm 0.53$ & $0.070 \pm 0.002$ & $0.86 \pm 0.04$ & $0.94 \pm 0.14$ & $5225 \pm 80$ & $+0.44 \pm 0.10$ & 10\\
 Kepler-39b & 21.087 & $20.1 \pm 1.3$ & $1.24 \pm 0.10$ & $0.112 \pm 0.057$ & $1.29 \pm 0.07$ & $1.40 \pm 0.10$ & $6350 \pm 100$ & $+0.10 \pm 0.14$ & 11\\
 KOI-189b & 30.360 & $78.0 \pm 3.4$ & $1.00 \pm 0.02$ & $0.275 \pm 0.004$ & $0.76 \pm 0.05$ & $0.73 \pm 0.02$ & $4952 \pm 40$ & $-0.07 \pm 0.12$ & 12\\
 KOI-205b & 11.720 & $39.9 \pm 1.0$ & $0.81 \pm 0.02$ & $<0.031$ & $0.92 \pm 0.03$ & $0.84 \pm 0.02$ & $5237 \pm 60$ & $+0.14 \pm 0.12$ & 13\\
 KOI-415b & 166.788 & $62.1 \pm 2.7$ & $0.79 \pm 0.12$ & $0.689 \pm 0.001$ & $0.94 \pm 0.06$ & $1.15 \pm 0.15$ & $5810 \pm 80$ & $-0.24 \pm 0.11$ & 14\\
 EPIC 201702477b & 40.737 & $66.9 \pm 1.7$ & $0.76 \pm 0.07$ & $0.228 \pm 0.003$ & $0.87 \pm 0.03$ & $0.90 \pm 0.06$ & $5517 \pm 70$ & $-0.16 \pm 0.05$ & 15\\
 EPIC 212036875b & 5.170 & $52.3 \pm 1.9$ & $0.87 \pm 0.02$ & $0.132 \pm 0.004$ & $1.29 \pm 0.07$ & $1.50 \pm 0.03$ & $6238 \pm 60$ & $+0.01 \pm 0.10$ &  18, 21 \\
 AD 3116b & 1.983 & $54.2 \pm 4.3$ & $1.02 \pm 0.28$ & $0.146 \pm 0.024$ & $0.28 \pm 0.02$ & $0.29 \pm 0.08$ & $3200 \pm 200$ & $+0.16 \pm 0.10$ & 17 \\
 CWW 89Ab & 5.293 & $39.2 \pm 1.1$ & $0.94 \pm 0.02$ & $0.189 \pm 0.002$ & $1.10 \pm 0.05$ & $1.03 \pm 0.02$ & $5755 \pm 49$ & $+0.20 \pm 0.09$ & 16, 18 \\
 RIK 72b & 97.760 & $59.2 \pm 6.8$ & $3.10 \pm 0.31$ & $0.146 \pm 0.012$ & $0.44 \pm 0.04$ & $0.96 \pm 0.10$ & $3349 \pm 142$ & - & 19\\
 TOI-503b & 3.677 & $53.7 \pm 1.2$ & $1.34^{+0.26}_{-0.15}$ & 0 (adopted) & $1.80 \pm 0.06$ & $1.70 \pm 0.05$ & $7650 \pm 160$ & $+0.61 \pm 0.07$ & 22 \\
 TOI-569b & 6.556 & $63.8 \pm 1.0$ & $0.75 \pm 0.02$ & $<0.0035$ & $1.21 \pm 0.03$ & $1.48 \pm 0.03$ & $5705 \pm 76$ & $+0.40 \pm 0.08$ & 24\\ 
 TOI-1406b & 10.574 & $46.0\pm 2.7$ & $0.86\pm 0.03$ & $<0.039$ & $1.18 \pm 0.09$ & $1.35\pm 0.03$ & $6290\pm 100$ & $-0.08 \pm 0.09$ & 24\\
 NGTS-7Ab$^\ast$ & 0.676 & $75.5^{+3.0}_{-13.7}$ & $1.38^{+0.13}_{-0.14}$ & 0 (adopted) & $0.48\pm 0.13$ & $0.61\pm 0.06$ & $3359\pm 106$ & - & 23\\
 2M0535-05a & 9.779 & $56.7 \pm 4.8$ & $6.50 \pm 0.33$ & $0.323 \pm 0.006$ & - & - & - & - & 20\\
 2M0535-05b & 9.779 & $35.6 \pm 2.8$ & $5.00 \pm 0.25$ & $0.323 \pm 0.006$ & - & - & - & - & 20\\
 2M1510 Aa & 20.902 & $40.0 \pm 2.9$ & $1.53 \pm 0.15$ & $0.309 \pm 0.022$ & - & - & - & - & 25\\
 2M1510 Ab & 20.902 & $39.9 \pm 2.9$ & $1.53 \pm 0.15$ & $0.309 \pm 0.022$ & - & - & - & - & 25\\
 \enddata
 \tablecomments{References: 1 - \cite{zhou19}, 2 - \cite{kelt1b}, 3 - \cite{irwin10}, 4 -  \cite{johnson11_bd}, 5 - \cite{irwin18}, 6 - \cite{wasp30b}, 7 - \cite{wasp128b}, 8 - \cite{corot3b}, 9 - \cite{corot15b}, 10 - \cite{corot33b}, 11 - \cite{kepler39}, 12 - \cite{diaz14}, 13 - \cite{diaz13}, 14 - \cite{moutou13}, 15 - \cite{bayliss16}, 16 - \cite{nowak17}, 17 - \cite{ad3116}, 18 - \cite{carmichael19}, 19 - \cite{david19_bd}, 20 - \cite{2M0535} (eclipsing system), 21 - \cite{persson2019}, 22 - \cite{subjak2019}, 23 - \cite{jackman2019}, 24 - \cite{carmichael20}, 25 - \cite{bd_triple} (eclipsing system), $^\ast$TESS aided in the discovery of NGTS-7Ab (TIC 2055733261)}

\end{deluxetable}

\begin{deluxetable*}{lcccc}
\tablecaption{MIST median values and 68\% confidence interval for TOI-811, created using {\tt EXOFASTv2} commit number  \dbf{f8f3437}. Here, $\mathcal{U}$[a,b] is the uniform prior bounded between $a$ and $b$, and $\mathcal{G}[a,b]$ is a Gaussian prior of mean $a$ and width $b$. We show $v\sin{I_\star}$ (measured from TRES) and the stellar age from Equation \ref{eq:gyro} ($Age_{\rm gyro}$) only for convenient reference; the current build of {\tt EXOFASTv2} does not model $v\sin{I_\star}$ or $Age_{\rm gyro}$. We do use $Age_{\rm gyro}$ as a prior on the age.}
\tablehead{\colhead{~~~Parameter} & \colhead{Units} & \colhead{Priors} & \multicolumn{2}{c}{Values}}
\startdata
\multicolumn{2}{l}{\bf Stellar Parameters:}&\smallskip\\
~~~~$M_*$\dotfill &Mass (\msun)\dotfill & - &$1.323^{+0.053}_{-0.069}$\\
~~~~$R_*$\dotfill &Radius (\rsun)\dotfill & - &$1.270^{+0.064}_{-0.092}$\\
~~~~$L_*$\dotfill &Luminosity (\lsun)\dotfill & - &$2.04^{+0.21}_{-0.32}$\\
~~~~$\rho_*$\dotfill &Density (cgs)\dotfill & - &$0.911^{+0.18}_{-0.099}$\\
~~~~$\log{g}$\dotfill &Surface gravity (cgs)\dotfill & - &$4.352^{+0.045}_{-0.029}$\\
~~~~$T_{\rm eff}$\dotfill &Effective Temperature (K)\dotfill & $\mathcal{G}[6013,56]$ &$6107\pm 77$\\
~~~~$[{\rm Fe/H}]$\dotfill &Metallicity (dex)\dotfill & $\mathcal{G}[0.40,0.10]$ &$0.402^{+0.065}_{-0.090}$\\
~~~~$Age$\dotfill &Age (Gyr)\dotfill & $\mathcal{G}[0.077,0.064]$ &$0.117^{+0.043}_{-0.037}$\\
~~~~$Age_{\rm gyro}$\dotfill &Age from gyrochronology (Gyr)\dotfill & Not modelled &$0.077^{+0.064}_{-0.026}$\\
~~~~$A_V$\dotfill &V-band extinction (mag)\dotfill & $\mathcal{U}[0,0.07976]$ &$0.055^{+0.040}_{-0.037}$\\
~~~~$\sigma_{SED}$\dotfill &SED photometry error scaling \dotfill & - &$8.9^{+5.0}_{-2.5}$\\
~~~~$\varpi$\dotfill &Parallax (mas)\dotfill & $\mathcal{G}[3.4923,0.0211]$ &$3.494^{+0.021}_{-0.022}$\\
~~~~$d$\dotfill &Distance (pc)\dotfill & - & $286.2^{+1.8}_{-1.7}$\\
~~~~$v\sin{I_\star}$\dotfill &Projected equatorial velocity ($\rm km\, s^{-1}$)\dotfill & Not modelled & $7.11 \pm 0.50$\\
~~~~$V_{\rm rot}$\dotfill &Rotational velocity ($\rm km\, s^{-1}$)\dotfill & Not modelled & $19.44^{+1.35}_{-2.44}$\\
~~~~$P_{\rm rot}$\dotfill &Rotation period (days)\dotfill & Not modelled & $3.21 \pm 0.02$\\
\multicolumn{2}{l}{\bf Brown Dwarf Parameters:}& \\
~~~~$P$\dotfill &Period (days)\dotfill & - &$25.16551^{+0.000035}_{-0.000033}$\\
~~~~$M_P$\dotfill &Mass (\mj)\dotfill  & - &$59.9^{+13}_{-8.6}$\\
~~~~$R_P$\dotfill &Radius (\rj)\dotfill & - & $1.262\pm0.062$\\
~~~~$T_C$\dotfill &Time of conjunction (\bjdtdb)\dotfill & - &$2458438.24914^{+0.00049}_{-0.00047}$\\
~~~~$a$\dotfill &Semi-major axis (AU)\dotfill  & - &$0.1874^{+0.0026}_{-0.0035}$\\
~~~~$i$\dotfill &Orbital inclination (Degrees)\dotfill & - &$89.27^{+0.44}_{-0.29}$\\
~~~~$e$\dotfill &Eccentricity \dotfill  & - &$0.509\pm0.075$\\
~~~~$ecos{\omega_*}$\dotfill & \dotfill & - &$-0.454^{+0.085}_{-0.091}$\\
~~~~$esin{\omega_*}$\dotfill & \dotfill & - &$0.218^{+0.041}_{-0.042}$\\
~~~~$T_{eq}$\dotfill &Equilibrium temperature (K)\dotfill  & - &$762^{+15}_{-16}$\\
~~~~$K$\dotfill &RV semi-amplitude ($\rm m\, s^{-1}$)\dotfill  & - &$3880^{+1000}_{-610}$\\
~~~~$R_P/R_*$\dotfill &Radius of planet in stellar radii \dotfill  & - &$0.1018^{+0.0012}_{-0.0011}$\\
~~~~$a/R_*$\dotfill &Semi-major axis in stellar radii \dotfill  & - &$31.6^{+1.3}_{-1.2}$\\
~~~~$\delta$\dotfill &Transit depth (fraction)\dotfill  & - &$0.01037^{+0.00024}_{-0.00021}$\\
~~~~$\tau$\dotfill &Ingress/egress transit duration (days)\dotfill  & - &$0.0189^{+0.0018}_{-0.0014}$\\
~~~~$b$\dotfill &Transit Impact parameter \dotfill  & - &$0.29^{+0.11}_{-0.16}$\\
~~~~$logg_P$\dotfill &Surface gravity \dotfill & -  &$4.976^{+0.066}_{-0.067}$\\
~~~~$M_P\sin i$\dotfill &Minimum mass (\mj)\dotfill & -  &$59.9^{+13}_{-8.6}$\\
~~~~$M_P/M_*$\dotfill &Mass ratio \dotfill & -  &$0.0434^{+0.0084}_{-0.0058}$\\
\hline
\\\multicolumn{2}{l}{Wavelength Parameters:}&I&TESS\\
~~~~$u_{1}$\dotfill &linear limb-darkening coeff \dotfill &$0.175\pm0.038$&$0.312^{+0.039}_{-0.040}$\\
~~~~$u_{2}$\dotfill &quadratic limb-darkening coeff \dotfill &$0.261^{+0.048}_{-0.047}$&$0.315\pm0.046$\\
~~~~$A_D$\dotfill &Dilution from neighboring stars \dotfill &$0.1920^{+0.0057}_{-0.0075}$&$0.1883^{+0.0074}_{-0.0059}$\\
\smallskip\\\multicolumn{2}{l}{Telescope Parameters:} & CHIRON \smallskip\\
~~~~$\gamma_{\rm rel}$\dotfill &Relative RV Offset (m/s)\dotfill &$34210^{+220}_{-240}$\\
~~~~$\sigma_J$\dotfill &RV Jitter (m/s)\dotfill &$620^{+210}_{-130}$\\
~~~~$\sigma_J^2$\dotfill &RV Jitter Variance \dotfill &$380000^{+300000}_{-150000}$\\
\smallskip\\\multicolumn{2}{l}{Transit Parameters:}& MEarth (I)& TESS  \smallskip\\
~~~~$\sigma^{2}$\dotfill &Added Variance \dotfill &$0.00000320^{+0.00000028}_{-0.00000027}$&$0.0000001919^{+0.0000000063}_{-0.0000000060}$\\
~~~~$F_0$\dotfill &Baseline flux \dotfill &$1.00003\pm0.00013$&$1.0000026^{+0.0000097}_{-0.0000098}$\\
\enddata
\end{deluxetable*}\label{tab:exofast_toi811}

\begin{deluxetable*}{lcccccc}
\tablecaption{MIST median values and 68\% confidence interval for TOI-852, created using {\tt EXOFASTv2} commit number \dbf{f8f3437}. The most likely values (probability of 0.58) and the ones we report for this system are shown in boldface. Here, $\mathcal{U}$[a,b] is the uniform prior bounded between $a$ and $b$, and $\mathcal{G}[a,b]$ is a Gaussian prior of mean $a$ and width $b$. We show $v\sin{I_\star}$ (measured from TRES) only for convenient reference; the current build of {\tt EXOFASTv2} does not model $v\sin{I_\star}$.}
\tablehead{\colhead{~~~Parameter} & \colhead{Units} & \colhead{Priors} & \colhead{\bf Most likely values} & \multicolumn{2}{c}{Less likely values}}
\startdata
\multicolumn{2}{l}{\bf Stellar Parameters:}& & $\rm \bf Prob.=0.58$& $\rm Prob.=0.42$\smallskip\\
~~~~$M_*$\dotfill &Mass (\msun)\dotfill & - &$\bf 1.32^{+0.05}_{-0.04}$ & $1.15^{+0.04}_{-0.05}$\\
~~~~$R_*$\dotfill &Radius (\rsun)\dotfill & - &$\bf 1.71\pm 0.04$ & $1.72\pm 0.04$ \\
~~~~$L_*$\dotfill &Luminosity (\lsun)\dotfill  & - &$\bf 2.92^{+0.17}_{-0.16}$ & $2.85^{+0.16}_{-0.15}$\\
~~~~$\rho_*$\dotfill &Density (cgs)\dotfill  & - &$\bf 0.37^{+0.03}_{-0.02}$ &$0.32\pm 0.02$\\
~~~~$\log{g}$\dotfill &Surface gravity (cgs)\dotfill  & - & $\bf 4.09\pm 0.04$ & $4.02^{+0.02}_{-0.03}$\\
~~~~$T_{\rm eff}$\dotfill &Effective Temperature (K)\dotfill  & $\mathcal{G}[5749,51]$ &$\bf 5768^{+84}_{-81}$ & $5714\pm 82$ \\
~~~~$[{\rm Fe/H}]$\dotfill &Metallicity (dex)\dotfill  & $\mathcal{G}[0.30,0.10]$ &$\bf +0.334^{+0.085}_{-0.088}$ &$+0.279^{+0.095}_{-0.094}$\\
~~~~$Age$\dotfill &Age (Gyr)\dotfill  & - &$\bf 4.04^{+0.68}_{-0.76}$& $7.29^{+1.3}_{-0.92}$\\
~~~~$A_V$\dotfill &V-band extinction (mag)\dotfill  & $\mathcal{U}[0,0.07471]$ &$\bf 0.038^{+0.025}_{-0.026}$&$0.037^{+0.026}_{-0.025}$\\
~~~~$\sigma_{SED}$\dotfill &SED photometry error scaling \dotfill  & - &$\bf 2.51^{+1.0}_{-0.62}$&$2.52^{+1.0}_{-0.61}$\\
~~~~$\varpi$\dotfill &Parallax (mas)\dotfill  & $\mathcal{G}[2.8172,0.0421]$ &$\bf 2.820^{+0.042}_{-0.043}$&$2.818^{+0.044}_{-0.042}$\\
~~~~$d$\dotfill &Distance (pc)\dotfill  & - &$\bf 354.7^{+5.5}_{-5.2}$&$354.9^{+5.3}_{-5.4}$\\
~~~~$v\sin{I_\star}$\dotfill &Projected equatorial velocity ($\rm km\, s^{-1}$)\dotfill & Not modelled & $\bf 14.50 \pm 0.5$ & $14.50 \pm 0.50$\\
~~~~$V_{\rm rot}$\dotfill &Rotational velocity ($\rm km\, s^{-1}$)\dotfill & Not modelled & $\bf 14.97\pm 0.34$ & $14.97\pm 0.34$\\
~~~~$P_{\rm rot}$\dotfill &Rotation period (days)\dotfill & Not modelled & $\bf 5.80 \pm 1.39$ & $5.80 \pm 1.39$\\
\multicolumn{2}{l}{\bf Brown Dwarf Parameters:}& &\smallskip\\
~~~~$P$\dotfill &Period (days)\dotfill & - &$\bf 4.945613^{+0.000083}_{-0.000076}$&$4.945611^{+0.000083}_{-0.000074}$\\
~~~~$M_P$\dotfill &Mass (\mj)\dotfill  & - &$\bf 53.7^{+1.4}_{-1.3}$ & $49.2^{+1.1}_{-1.4}$\\
~~~~$R_P$\dotfill &Radius (\rj)\dotfill & - &$\bf 0.829^{+0.037}_{-0.035}$ & $0.85\pm 0.04$\\
~~~~$T_C$\dotfill &Time of conjunction (\bjdtdb)\dotfill & - &$\bf 2458387.1888^{+0.0020}_{-0.0019}$ & $2458387.1889^{+0.0021}_{-0.0020}$\\
~~~~$a$\dotfill &Semi-major axis (AU)\dotfill  & - &$\bf 0.06300^{+0.00077}_{-0.00073}$ & $0.06026^{+0.00071}_{-0.00089}$\\
~~~~$i$\dotfill &Orbital inclination (Degrees)\dotfill & - &$\bf 84.35^{+0.30}_{-0.29}$ & $83.82^{+0.32}_{-0.31}$\\
~~~~$e$\dotfill &Eccentricity \dotfill  & - &$\bf 0.0036^{+0.0043}_{-0.0025}$ & $0.0035^{+0.0042}_{-0.0025}$\\
~~~~$ecos{\omega_*}$\dotfill & \dotfill & - & $\bf 0.0000^{+0.0028}_{-0.0026}$ & $0.0001^{+0.0027}_{-0.0025}$\\
~~~~$esin{\omega_*}$\dotfill & \dotfill & - & $\bf 0.0001^{+0.0043}_{-0.0035}$ & $0.0001^{+0.0041}_{-0.0035}$\\
~~~~$T_{eq}$\dotfill &Equilibrium temperature (K)\dotfill  & - &$\bf 1449^{+23}_{-22}$ & $1476\pm23$\\
~~~~$K$\dotfill &RV semi-amplitude ($\rm m\, s^{-1}$)\dotfill  & - &$\bf 5182^{+29}_{-32}$ & $5182^{+29}_{-32}$\\
~~~~$R_P/R_*$\dotfill &Radius of planet in stellar radii \dotfill & -  &$\bf 0.0498\pm0.0011$ & $0.0503\pm0.0012$\\
~~~~$a/R_*$\dotfill &Semi-major axis in stellar radii \dotfill  & - &$\bf 7.92^{+0.24}_{-0.23}$ & $7.49^{+0.24}_{-0.23}$\\
~~~~$\delta$\dotfill &Transit depth (fraction)\dotfill  & - &$\bf 0.00248\pm0.00011$ & $0.00253\pm0.00012$\\
~~~~$\tau$\dotfill &Ingress/egress transit duration (days)\dotfill  & - &$\bf 0.0160\pm0.0013$ & $0.0182\pm0.0015$\\
~~~~$b$\dotfill &Transit Impact parameter \dotfill  & - &$\bf 0.780^{+0.019}_{-0.023}$ & $0.806^{+0.017}_{-0.021}$\\
~~~~$logg_P$\dotfill &Surface gravity \dotfill  & - &$\bf 5.288\pm0.037$ & $5.229\pm0.038$\\
~~~~$M_P\sin i$\dotfill &Minimum mass (\mj)\dotfill  & - &$\bf 53.4^{+1.4}_{-1.3}$ & $48.9^{+1.2}_{-1.5}$\\
~~~~$M_P/M_*$\dotfill &Mass ratio \dotfill  & - &$\bf 0.03903^{+0.00053}_{-0.00055}$ & $0.04094^{+0.00068}_{-0.00056}$\\
\hline
\\\multicolumn{2}{l}{Wavelength Parameters:}&R&TESS\smallskip\\
~~~~$u_{1}$\dotfill &linear limb-darkening coeff \dotfill &$0.388^{+0.052}_{-0.053}$&$0.317\pm0.052$\\
~~~~$u_{2}$\dotfill &quadratic limb-darkening coeff \dotfill &$0.275\pm0.050$&$0.279^{+0.050}_{-0.049}$\\
\smallskip\\\multicolumn{2}{l}{Telescope Parameters:}&CORALIE&TRES\smallskip\\
~~~~$\gamma_{\rm rel}$\dotfill &Relative RV Offset (m/s)\dotfill &$-16896^{+32}_{-38}$&$-333\pm34$\\
~~~~$\sigma_J$\dotfill &RV Jitter (m/s)\dotfill &$60\pm60$&$64^{+48}_{-61}$\\
~~~~$\sigma_J^2$\dotfill &RV Jitter Variance \dotfill &$3600^{+11000}_{-4900}$&$4100^{+8500}_{-4100}$\\
\smallskip\\\multicolumn{2}{l}{Transit Parameters:}&ULMT (R) & TESS \smallskip\\
~~~~$\sigma^{2}$\dotfill &Added Variance \dotfill &$0.00000145^{+0.00000037}_{-0.00000029}$&$0.0000001869^{+0.0000000088}_{-0.0000000082}$\\
~~~~$F_0$\dotfill &Baseline flux \dotfill &$1.00025\pm0.00018$&$0.999997\pm0.000014$\\
\enddata
\end{deluxetable*}\label{tab:exofast_toi852}

\end{document}